\documentclass[11pt]{article}
\usepackage{amsmath,amssymb}
\usepackage[dvips]{graphicx}
\usepackage[nosort]{cite}

\begin{document}

\baselineskip=17pt

\begin{titlepage}
\rightline{\tt arXiv:1211.2649}
\rightline{\tt UT-Komaba/12-11}
\begin{center}
\vskip 2.5cm
{\Large \bf {Comments on new multiple-brane solutions based on}}\\
\vskip 0.3cm
{\Large \bf {Hata-Kojita duality in open string field theory}}\\
\vskip 0.75cm
{\large {Toru Masuda}}
\vskip 0.5 cm
{\it {Institute of Physics, The University of Tokyo,}}\\
{\it {Komaba, Meguro-ku, Tokyo 153-8902, Japan}}\\

\vskip 0.4cm

{\it {Department of Physics, Ochanomizu University,}}\\
{\it {Otsuka, Bunkyo-ku, Tokyo 2-1-1, Japan}}\\

\vskip 0.4cm

{E-mail: \tt masudatoru@gmail.com}

\vskip 1.64cm

{\bf Abstract}
\end{center}

\noindent
Recently, Hata and Kojita 
proposed a new energy formula 
for a class of solutions 
in Witten's open string field theory
based on a novel symmetry 
of correlation functions
they found. 
Their energy formula  
can be regarded as a generalization of 
the conventional energy formula by Murata and Schnabl. 
Following their proposal, we investigate their new ansatz for the  classical solution representing double D-branes.   
We present a regularized 
definition of this solution
and show that the solution satisfies the equation of motion when it is contracted with
the solution itself and when it is contracted with any states in the Fock space.
However, the Ellwood invariant  and 
the boundary state of the solution are the same as those for the perturbative vacuum.  
This result 
disagrees with an expectation from the Ellwood conjecture.

\end{titlepage}

\newpage

\tableofcontents

\section{Introduction}

Since the seminal study of Murata and Schnabl 
\cite{Murata:2011ex,Murata:2011ep},
 solutions for multiple D-branes
 in Witten's open string field theory 
 \cite{Witten:1985cc}
have been intensively considered. 
Very recently, there has appeared an interesting paper by Hata and Kojita\cite{Hata:2012cy}.
They proposed a new way to construct multiple-brane 
solutions. 
Their work can potentially reform our conventional understanding on this subject.

The starting point for the discussion is 
estimation of  the energy density of the Okawa-type solution, 
\begin{equation}\label{Okawa-solution}
\Psi_F=F(K)^2 c\frac{KB}{1-F(K)^2}c\,.
\end{equation}
Here $K$, $B$ and $c$ are symbols introduced to conveniently express a class of wedge states with operator 
insertions\cite{Okawa:2006vm}. 
These three symbols satisfy simple
algebraic relations called the $KBc$ subalgebra. 
We will review them in $\S$ 2.1. 
It is natural to expect that
physical properties of the solution $\Psi_F$ are determined by choice of
$F(K)$, which is a function of $K$. 
Murata and Schnabl derived 
a formula for the energy density of the solution
\eqref{Okawa-solution} under some holomorphy conditions  on $G(K)=1-F(K)^2$, 
\begin{equation}\label{Murata-Schnabl energy}
\mathcal E(\Psi_F)\sim\frac{n_0}{2\pi^2g_o^2}+(\text{anomalous term})\,.
\end{equation}
Here $n_a$ denotes the order of the pole (or the multiplicity of the zero times minus one) of $G(z)$ at $z = a$.
If we admit a formal object $1/K^n$ and drop some surface terms, 
the anomalous term does not appear; 
however, it is not zero in the calculation of \cite{Murata:2011ep}, except for the case $n_0=\pm1, 0$.
Despite several efforts, fully acceptable definition of multiple-brane solutions based on \eqref{Murata-Schnabl energy} is 
 not yet obtained.\footnote{In appendix C, we summarize our 
attempt to construct the double-brane solution based on \eqref{Murata-Schnabl energy}.  Although it is not successful, 
the regularization method obtained there is essential in the present work. }

In \cite{Hata:2012cy}, Hata and Kojita argued that 
the  pole at $K=-\infty$ is in a sense 
equivalent to 
the pole at $K=0$, and it also contribute to the energy density. 
This argument arise from a novel symmetry of 
correlation functions in the $KBc$ subspace.
This observation, together with \eqref{Murata-Schnabl energy}, 
lead us to the following energy formula\footnote
{
To be precise, the discussion of \cite{Hata:2012cy} is based on    
 a particular regularization scheme, 
and 
the explicit form of the anomalous term is also derived.}
\begin{equation}\label{HK_energy1}
\mathcal E(\Psi_F)\sim\frac{1}{2\pi^2g_o^2}(n_0+n_{-\infty})+(\text{anomalous term})\,.
\end{equation} 
From this formula, we can make new ansatzes for multiple-brane solutions.  

We here note that some previous studies 
\cite{Ellwood:2008jh,KOZ,Takahashi:2011wk,Masuda:2012kt,Erler:2012dz} do not appear to be  
consistent with \eqref{HK_energy1}. 
Still, it might be premature to dismiss \eqref{HK_energy1}. 
The pole at $K=-\infty$ is not fully considered so far, 
while it is related to the singularity of the identity string field.  
We empirically know that the identity string field requires quite careful treatment. 
Further investigation might resolve these apparent conflicts.

In this paper, we 
study 
 the following ansatz of a classical solution presented in \cite{Hata:2012cy}:\footnote{Notation used in \cite{Hata:2012cy}
 is different 
 from ours.  
 See appendix D for details.   }
\begin{equation}\label{new-double-brane-solution}
\Psi=Kc\frac{K}{1-K}Bc\,.
\end{equation}
This expression is singular 
in the sense that 
the energy density of \eqref{new-double-brane-solution} is indefinite. 
We then define a double-brane solution as a limit of sequence of regular string fields as follows: 
\begin{equation}\label{sln1}
\Psi=\lim_{\epsilon\to 0}K1_{\epsilon}c\frac{K}{1-K}Bc\,.
\end{equation}
Here $1_{\epsilon}$ is a regularized identity string field, defined by
\begin{equation}\label{1ee}
1_{\epsilon}=\int_0^\infty \delta_{\epsilon}(x)e^{xK}\,.
\end{equation}
The $\delta$-sequence on the right-hand side of \eqref{1ee} has
some special property, which is essential for our calculation. We will describe it in $\S$ 3.1.
As an example of $\delta_\epsilon(x)$, we can take the following:
\begin{equation}
\delta_\epsilon(x)=
\frac{1}{\log\left(\epsilon^{-1}+1\right)}
\frac{1}{(x+\epsilon)(x+\epsilon+1)}
\,.
\end{equation}
We summarize properties of the solution \eqref{sln1} below:
\begin{enumerate}
\item[(1)] The solution satisfies the equation of motion  when it is contracted with any state in the Fock space.
\item[(2)] The solution satisfies the equation of motion when it is contracted 
to the solution itself.
\item[(3)] The solution reproduces the energy density for double D-branes.
\end{enumerate}

This is the first multiple-brane solution which satisfies both (1) and (2);  
however,  
 it is still not clear whether (1) and (2) are sufficient to ensure that $\Psi$ is a solution to the equation of motion, since we do not know any good definition of the state space of the open string field theory.\footnote{
In this paper, we use the word {\it solution} to refer to $\Psi$ in \eqref{sln1} for simplicity; however, it is not precise in this sense. }
Indeed,
as we will see in $\S$ 4.4 and $\S$ 4.5, some existing conjectures contradict with the 
interpretation that \eqref{sln1} 
is a classical solution representing 
double D-branes. 
The property (3) is consistent with the formula \eqref{HK_energy1}. 
Since our calculation is completely independent of the argument of \cite{Hata:2012cy}, this agreement is interesting.

We note that regularization of the multi-brane solution is also claimed in \cite{Hata:2012cy} ($K_{\epsilon\eta}$ regularization). 
Yet, if one calculates the energy density of \eqref{new-double-brane-solution}
under the $K_{\epsilon\eta}$ regularization, one needs to use an  analytic continuation method called 
the $s$-$z$ trick \cite{Murata:2011ex} to obtain a finite value. 
Indeed, without the $s$-$z$ trick, one needs to drop singular terms by hand. 
This fact is explained in $\S$ 2.5 of \cite{Hata:2012cy}.  
This means that 
if one uses the $s$-$z$ trick, then the result is not equal to the original expression 
in some cases.  
Since the $s$-$z$ trick drastically simplifies calculation in many cases, and 
it is used in several studies, it is important to clarify that when 
it can be used as an identical transformation and when it cannot. 

This paper is organized as follows: 
In $\S$ 2, we briefly introduce  some preliminary materials and sketch a derivation of the formula \eqref{HK_energy1}.  
In $\S$ 3, 
we introduce the delta sequences appearing in \eqref{1ee}, which is essential for our regularization. 
In $\S$ 4, we present the definition of the double-brane solution and check the equation of motion; and then we also calculate some physical quantities including the energy density, the Ellwood invariant and the boundary state.  
In $\S$ 5, we summarize our results.

\section{Review}
\setcounter{equation}{0}
\subsection{The $KBc$ algebra}

In 2005, Schnabl constructed an analytic solution for tachyon condensation in 
Witten's open string field theory,
using 
a class of wedge states with operator insertions\cite{Schnabl:2005gv}. 
The $KBc$ subalgebra was introduced by Okawa to express this class of wedge states with operator insertions\cite{Okawa:2006vm}. 
Here $K$ is a grassman even object, and the wedge state $|n+1\rangle $ is represented as $e^{nK}$\,.
The object $B$ is grassman odd, and it represents line integral of the anti-ghost. 
The object $c$ is also grassman odd, and it represents insertion of the $c$-ghost at the boundary.  
Together with the usual BRST operator $Q$,
they satisfy the following algebraic relations:
\begin{equation}\label{KBc1}
[\,K,\,B\,]=0\,,\quad \{\,B,\,c\,\}=1\,,\quad c^2=B^2=0\,,
\end{equation}
\begin{equation}\label{KBc2}
Q K=0\,,\quad Q B=K\,,\quad Qc=cKc\,.
\end{equation}
As described by Erler in \cite{Erler:2006hw}, we can regard $K$, $B$ and $c$ as identity-based string fields. 
The commutation relations \eqref{KBc1} then can be written using the star product: $[K,\,B]=K\ast B-B\ast K$, etc.\footnote{The product symbol $\ast$ is usually omitted when we express string fields using $K$, $B$ and $c$. }. 
The space of string fields which can be written by $K$, $B$ and $c$ is closed under 
the star multiplication and the action of $Q$. 
Thus we can use the $KBc$ subalgebra to find solutions to the equation of motion. 

Consider the following formal solution to the equation of motion:
\begin{equation}\label{Okawa}\Psi_F=F(K)^2c\frac{KB}{1-F(K)^2}c\,.\end{equation}
Here $F(K)$ is a function of $K$. Choice of $F(K)$ determines physical properties of
the solution. 
For instance, the choice $F(K)=e^{K/2}$ corresponds to Schnabl's original 
tachyon vacuum solution, while the choice $F(K)=(1-K)^{-1/2}$ corresponds to 
the simple tachyon vacuum solution by Erler and Schnabl\cite{Erler:2009uj}. 
Each of these two solutions reproduces the energy density of the tachyon vacuum.  
We can also construct solutions with zero energy density with a suitable choice of $F(K)$.

It is remarkable that the Okawa-type solution \eqref{Okawa} can formally be written as a pure-gauge form:
\begin{equation}\label{pgf}
\Psi_F=UQU^{-1}\,,
\end{equation}
where 
\[U=1-F(K)^2Bc\,,\qquad
\text{and}\qquad
U^{-1}=1+\frac{F(K)^2}{1-F(K)^2}Bc\,.\]
Since $U$ or $U^{-1}$ might be singular in general,
 $\Psi_F$ is not necessarily be a pure-gauge solution\,. 
For example, if we take $F(K)=(1-K)^{-1/2}$, 
which corresponds to the simple tachyon vacuum solution, 
then the string field $U$ has a factor $1/K$. 
Generalizing the Okawa-type solution, a  class of formal solutions which can  formally be written as a pure-gauge form is presented in \cite{Masuda:2012kt}. 

We mostly follow the convention of \cite{Okawa:2006vm}, except for the overall factors of $K$, $B$ and $c$. 
See appendix D for details.   
For introduction to these topics including   the $KBc$ subalgebra, see reviews\cite{ReviewSchnabl,ReviewOkawa}. 

\subsection{The inversion symmetry}
In this subsection, we summarize a derivation of the formula \eqref{HK_energy1}. 
Let us start from the following homomorphisms of the $KBc$ subalgebra
\cite{Erler:2012dz,Masuda:2012kt},
\begin{equation}\label{HK_transf}
\hat K=f(K)\,,\quad \hat B=\frac{f(K)}{K}B\,,
\quad \hat c=c\frac{K}{f(K)}Bc\,.
\end{equation}
These hatted objects, $\hat K,\,\hat B$ and $\hat c$, 
satisfy the same algebraic relations as the original 
$K,\, B$ and $c$ :
\begin{equation}
[\, \hat K,\,\hat B\,]=0\,,\quad \{\,\hat B,\, \hat c\, \}=1\,,\quad \hat c^2=\hat B^2=0\,,
\end{equation}
\begin{equation}
Q \hat K=0\,,\quad Q \hat B=\hat K\,,\quad Q\hat c=\hat c\hat K\hat c\,.
\end{equation}
Transformation law of the Okawa-type solution under these homomorphisms is simple,
\begin{equation}\label{trl}
F(\hat K)^2\hat c\frac{\hat K\hat B}{1-F(\hat K)^2}\hat c=F(f(K))^2 c\frac{KB}{1-F(f(K))^2}c\,.
\end{equation}
That is, $\hat \Psi_{F}=\Psi_{\hat F}$ .

Let us concentrate on the special case $f(K)=1/K$. We define $\tilde K$, $\tilde B$ and $\tilde c$ as 
\begin{equation}\label{HK_transf}
\tilde K=\frac{1}{K}\,,\quad \tilde B=\frac{1}{K^2}B\,,
\quad \tilde c=cK^2Bc\,.
\end{equation}
Hata and Kojita proved the following symmetry of the correlation function (the inversion symmetry):
\begin{equation}\label{mvl}
{\rm tr}[\tilde B \tilde ce^{x_1\tilde K}
\tilde ce^{x_2\tilde K}\tilde ce^{x_3\tilde K}
\tilde ce^{x_4\tilde K}]\cong
{\rm tr}[Bce^{x_1 K}ce^{x_2 K}ce^{x_3 K}ce^{x_4 K}]\,.
\end{equation}
Note that in principle all the correlation functions in the $KBc$ subalgebra can be written using the four-point function on the right-hand side of \eqref{mvl}; 
given a correlation function in the $KBc$ subalgebra, one can reduce the number of insertions of $B$  using the anti-commutation relation of $B$ and $c$. Insertions of $K$s can be replaced with multiple differentiation of $x_i$\,. 
Thus, correlation functions in the $KBc$ subalgebra are invariant under the replacement of ($K$,\ $B$, $c$) by ($\tilde K$,\ $\tilde B$, $\tilde c$)  in general. 

Now, suppose that we 
find the energy density $\mathcal E(\Psi_F)$ of the Okawa-type solution 
$\Psi_F$ for a choice of $F(K)$. 
If we replace ($K$, $B$, $c$) 
 in this calculation of $\mathcal E(\Psi_F)$  by 
($\tilde K$, $\tilde B$, $\tilde c$),  
the resulting value  does not change, because of  the inversion symmetry \eqref{mvl}; while the solution $\Psi_{F(K)}$ 
becomes
$\Psi_{\tilde F(K)}= \Psi_{F(1/K)}$, from \eqref{trl}, under this replacement. 
Therefore, %
the energy density of the solution $\Psi_{F}$ is the same as that of 
$\Psi_{F(1/K)}$\,,
\begin{equation}\label{energy-duality}
\mathcal E(\Psi_{F(K)})=\mathcal E(\Psi_{F(1/K)})\,.
\end{equation}
We call this relation the Hata-Kojita duality. 
From \eqref{energy-duality} and \eqref{Murata-Schnabl energy}, we are lead to~\eqref{HK_energy1}.\footnote{Note that the pole of the function $G(K)=1-F(K)^2$ at $K=-\infty$ is not allowed in \eqref{Murata-Schnabl energy}. 
}

Note that we used the symbol $\cong$ rather than $=$ in \eqref{mvl}. The reason is that 
the left-hand side has some singular terms, and the value is indefinite in the usual sense; if we define the string field $e^{{x}/{K}}$ as 
\[e^{{x}/{K}}=1- \int_0^\infty x \, _0{F}_1(;2;-t x)e^{tK}dt\,, \]
where $_0{F}_1(;a;z)$ denotes a confluent hypergeometric function, 
\[_0{F}_1(;a;z)=
\sum_{k=0}^\infty\frac{1}{(a)_k}\frac{z^k}{k!}\,,\]
then 
the left-hand side of \eqref{mvl} contains some identity-based terms, such as 
${\rm tr}[BcK^2cK^2c^2K^2]$. 
To maintain the equivalence, we need to drop these terms. 
They are naturally dropped when we use the $s$-$z$ trick. 
\subsection{Defining the solution as a limit}\label{pop}
In this paper, we define the double-brane solution as a limit of a sequence of regular string fields (see \eqref{sln1}). That is, we consider one parameter family of string fields $\Psi_\epsilon$ 
with a small parameter $\epsilon>0$, and 
regard the limit $\Psi\equiv \lim_{\epsilon\to 0}\Psi_\epsilon$ as a solution to the equation
of motion. 
We would like to clarify this point in the following. 

For nonzero $\epsilon$, the string field $\Psi_\epsilon$ does not satisfy the equation of motion:
\begin{equation}
\text{eom}(\Psi_\epsilon)\equiv Q\Psi_\epsilon+\Psi_\epsilon\ast\Psi_\epsilon\ne 0\,.
\end{equation}
We would like to require that the contraction of $\text{eom}(\Psi_\epsilon)$ and any state $\varphi$ in the state space of the open string field theory vanishes as $\epsilon$ approaches  $0$:
\begin{equation}\notag
\lim_{\epsilon\to 0+}\langle\,  \varphi\, \big|\, \text{eom}(\Psi_\epsilon)\, \rangle=0\,.
\end{equation}
However, we do not know how to define the state space of the open string field theory. 
We then only require that  $\lim_{\epsilon\to 0}\langle\,  \varphi\, \big|\, \text{eom}(\Psi_\epsilon)\, \rangle$ vanishes when $\varphi$ is any state in the Fock space and 
when $\varphi$ is the solution $\lim_{\epsilon\to 0}\Psi_\epsilon$ itself. 
Note that there is no relationship between these two requirements in general, for the state $\lim_{\epsilon\to 0}\Psi_\epsilon$
 usually lies  outside the Fock space. 

When we calculate the physical quantities from the solution, we take the limit
$\epsilon\to 0$ at the end of the calculation. 
For example, the energy density of $\Psi$ is defined as follows:
\begin{equation}\notag
\mathcal E(\Psi)=\lim_{\epsilon\to 0}\frac{1}{g_o^2}
\left(\,
\frac{1}{2}{\rm tr} [\,\Psi_\epsilon\,Q\,\Psi_\epsilon]+
\frac{1}{3}{\rm tr} [\,\Psi_\epsilon\,\Psi_\epsilon\,\Psi_\epsilon\,]\,
\right)\,.
\end{equation}
We define the Ellwood invariant\footnote{It is also commonly referred to as the gauge-invariant observable or
the gauge-invariant overlap.} 
and the boundary state in a similar fashion.
See $\S$\ref{el} and $\S$\ref{bo} for details. 
Above treatment of the equation of motion and physical quantities 
reflect an expectation that the state space of the open string field theory is complete with respect to some norm.

\section{Regularization}
\setcounter{equation}{0}
In this section, we describe the regularization method
 used in this paper. 
In $\S$3.1, we describe our delta sequence $\delta_\epsilon(x)$\,. 
In $\S$3.2, we describe the regularized identity state $1_\epsilon$\,.

\subsection{A class of $\delta$-sequence}
Consider one parameter family of positive functions $\{\delta_\epsilon(x)\}$ with a 
small parameter $\epsilon>0$\,. 
We require the following conditions on $\delta_\epsilon(x)$:
\begin{enumerate}
\item $\lim_{\epsilon\to 0}\int_a^\infty\delta_{\epsilon}(x)dx=0 $ for $^\forall a>0$\,. 
\item $\int_0^\infty \delta_\epsilon(x)dx=1$ for $^\forall \epsilon>0$.
\end{enumerate}
Note that the lower limit of the integral in the condition 2 is zero. 
Since $\delta_\epsilon(x)$ is positive, $\int^x_0\delta_\epsilon(t)dt$ is a monotonically increasing function of $x$. 
Let $\lambda_\epsilon(t)$ be the inverse function of $\int^x_0\delta_\epsilon(t)dt$,
\[y=\int_0^x\delta_\epsilon(t)dt
\quad \longleftrightarrow \quad
x=\lambda_\epsilon (y)\,.\]
We further require the following 
special condition on $\lambda_\epsilon(t)$:
\begin{enumerate}
\item[$3.$]For $0<^\forall{a}<^\forall b<1$, 
\begin{equation}\notag
\lim_{\epsilon\to 0+}\frac{\lambda_{\epsilon}(a)}{\lambda_{\epsilon}(b)}=0\,.
\end{equation} 
\end{enumerate}
These three conditions characterize our delta sequence $\delta_\epsilon(t)$. 

Now, we would like to prove the following property of $\delta_\epsilon(x)$: 
If $f(s,\,t)$ is a bounded function on
$(s,\ t)\in D$, where 
$D= [0,\ \infty)\times[0,\ \infty)\backslash\{(0,\ 0)\}$,
then it follows that\footnote
{To be precise, 
we also assume that $f_0(x,\ y)\equiv\lim_{r\to 0+}f(xr,\ yr)$ ($(x,\ y)\in D$) is continuous at $(1,\,0)$ and
$(0,\,1)$. 
} 
\begin{equation}\label{trn}
\lim_{\epsilon\to 0}\int_0^\infty ds \int_0^\infty dt\, \delta_\epsilon(s) \delta_\epsilon(t)f(s,\,t)
=\frac{
\displaystyle \lim_{a\to 0+}f(a,\,0)+\lim_{a\to 0+}f(0,\,a)}{2}\,.
\end{equation}
To prove \eqref{trn}, we change the variables of integration on the left-hand side:
\begin{equation}\notag
\begin{split}
&\int_0^\infty ds_1 \int_0^\infty ds_2\, \delta_\epsilon(s_1)\, \delta_\epsilon(s_2)\,f(s_1,\,s_2)\\
=&\int_0^\infty ds_1 \int_0^\infty ds_2\, 
\frac{dh_\epsilon(s_1)}{ds_1} \frac{dh_\epsilon(s_2)}{ds_2}\,f(s_1,\,s_2)\\
=&\int_0^1dh_1 \int_0^1dh_2\, f(\lambda_\epsilon(h_1),\,\lambda_\epsilon(h_2))\,.
\end{split}
\end{equation}
Here, $h_\epsilon(s)$ in the second line denotes $h_\epsilon(s)=\int_0^s \delta_\epsilon(x)dx$.
Let us divide the integration region of $(h_1,\ h_2)$ into three parts:
\[T_1\equiv \{(h_1,\,h_2)\big|0\le h_1<h_2<1\}\,,\]
\[T_2\equiv \{(h_1,\,h_2)\big|0\le h_2<h_1<1\}\,,\]
\[F\equiv\{(h_1,\,h_2)\big|0\le h_1,\,h_2\le 1, \ (h_1-h_2)(h_1-1)(h_2-1)=0\}\,.\]
If $(h_1,\ h_2)\in T_1$, 
then the ratio $\lambda_\epsilon(h_1)/\lambda_\epsilon(h_2)$ converges to 
$0$ as $\epsilon$ approaches 0.  From the conditions~1~and~2, we also see that  both  
$\lambda_\epsilon(h_1)$ 
and 
$\lambda_\epsilon(h_2)$  
converge to 0 as $\epsilon$ approaches 0. 
Thus, we see the following: 
\begin{equation}\label{conv1}
f(\lambda_\epsilon(h_1),\,\lambda_\epsilon(h_2))\to \lim_{a\to 0}f(0, a)\,,\qquad(h_1,\,h_2)\in T_1\,.
\end{equation}
We present a rigorous proof of \eqref{conv1} 
in appendix A. 
Similarly, assuming $(h_1,\,h_2)\in T_2$, 
the integrand $f(\lambda_\epsilon(h_1) \,\lambda_\epsilon(h_2))$ converges to 
$\lim_{a\to 0}f(a, 0)$ as $\epsilon$ approaches 0:  
\begin{equation}\label{conv2}
f(\lambda_\epsilon(h_1),\,\lambda_\epsilon(h_2))\to \lim_{a\to 0}f(a, 0)\,,\qquad(h_1,\,h_2)\in T_2\,.
\end{equation}
Since $f(s,\,t)$ is bounded, the integration over $F$ is zero. 
Therefore, we obtain \eqref{trn}. 
We can generalize \eqref{trn} to multi-variable integrations. 
For three variables, we can prove that\footnote{
We assume that 
$g(x,\,y,\,z)$ is a bounded function on 
$D^{(3)}=\{\,(x,\,y,\,z)\,|\,x,\,y,\,z\ge 0, \,(x,\,y,\,z)\ne (0,\,0,\,0)\}$, and 
$g_{0}(x,\,y,\,z)\equiv \lim_{r\to 0+}g(xr,\,yr,\,zr)$ 
is continuous at (0,\,0,\,1), (0,\,1,\,0) and (1,\,0,\,0). 
} 
\begin{equation}\label{3rln}
\begin{split}
&\lim_{\epsilon\to 0}
\int_0^\infty ds_1 \int_0^\infty ds_2\int_0^\infty ds_3\, 
\delta_\epsilon(s_1)\delta_\epsilon(s_2)\delta_\epsilon(s_3)\ 
g(s_1,\,s_2,\,s_3)\\
=&\frac{1}
{3} \lim_{a\to 0+}\bigg(
g(a,\,0,\,0)+g(0,\,a,\,0)+g(0,\,0,\,a)\bigg)\,.
\end{split}
\end{equation}

As a simple example of $\delta_{\epsilon}(x)$, we may take
\[\lambda_\epsilon(t)
=\frac{1}{\epsilon'^{t-1}-1}
-\frac{1}{\epsilon'^{-1}-1}\,,
\qquad 
\epsilon'=\frac{\epsilon}{1+\epsilon}
\,,\]
and 
\begin{equation}\label{example}
\delta_\epsilon(x)=
\frac{1}{\log\left(\epsilon^{-1}+1\right)}
\frac{1}{(x+\epsilon)(x+\epsilon+1)}
\,.
\end{equation}
Note that this choice satisfies the 
following stronger condition: 
\begin{enumerate}
\item[$3^*.$]For $0<a<b<1$ and $0<r$,
\begin{equation}\label{stronger}
\lim_{\epsilon\to 0}\frac{\lambda_{\epsilon}(a)}{\lambda_{r\epsilon}(b)}=0\,.
\end{equation} 
\end{enumerate}
Then, it follows that even if we change the 
ratio of two small parameters on the left-hand side 
of \eqref{trn}, 
the right-hand side does not 
change:
\begin{equation}\label{trn2}
\lim_{\epsilon\to 0}\int_0^\infty ds \int_0^\infty dt\ \delta_\epsilon(s) \,\delta_{r \epsilon}(t)\,f(s,\,t)
=\frac{
\displaystyle \lim_{a\to 0+}f(a,\,0)+\lim_{a\to 0+}f(0,\,a)}{2}\,,\qquad (0<r)\,.
\end{equation}

\subsection{Regularization of the identity state}
Using the delta sequence $\delta_\epsilon(x)$ in \eqref{example}, we define the regularized identity state $1_\epsilon$,
\[1_\epsilon=\int_0^\infty \delta_\epsilon(x)e^{xK}\,. \]
Some correlators in the $KBc$ subalgebra are  singular, and their singularity is related to the identity string field. %
We can use the object $1_\epsilon$ to regularize some of these correlators. 
For example, we regularize the correlator  
$\text{tr}[cKcKcK]$ as follows: 
\begin{equation}\label{regKcKcKc}
\begin{split}
&\lim_{\epsilon\to 0} \text{tr}\left[\,c1_\epsilon Kc1_\epsilon Kc1_\epsilon K\,\right]\\
=&\lim_{a\to 0} \text{tr}\left[\,ce^{aK} Kc KcK\,\right]\\
=&\lim_{x_1\to 0}\left(
\lim_{x_2\to 0}\lim_{x_3\to 0}
\frac{\partial}{\partial x_1}
\frac{\partial}{\partial x_2}
\frac{\partial}{\partial x_3}
{\rm tr}\left[\,ce^{x_1K}ce^{x_2K}ce^{x_3K}\,\right]
\right)\\
=&0\,.
\end{split}
\end{equation}
where we used \eqref{3rln}. 
An explicit form of the correlation function ${\rm tr}[ce^{x_1K}ce^{x_2K}ce^{x_3K}]$ is obtained by substituting $x_4=0$ in~\eqref{s6}. 
Assuming \eqref{stronger}, we can change the ratios of small parameters, 
\[
\lim_{\epsilon\to 0} \text{tr}[\,c1_\epsilon Kc1_{\alpha\epsilon} Kc1_{\beta\epsilon} K\,]=0,
\qquad \alpha,\,\beta>0\,.
\]

Taking the opportunity, we comment on the regularization of the identity-based solution in
the $KBc$ subalgebra, $\Psi=-(1+K)c$. We define a regularized solution as follows: 
\begin{equation}\label{id}
\Psi=-\lim_{\epsilon\to 0}(1+K)1_\epsilon c\,.
\end{equation}
From \eqref{regKcKcKc}, it follows that 
\[\text{tr}[\Psi\Psi\Psi]=0\,.\]
Similarly, we find
\[\text{tr}[\Psi Q\Psi]=0\,.\]
That is, $\Psi$ satisfies the equation of motion when it is contracted with the solution itself, and its energy density is zero.  
This result agrees with \eqref{HK_energy1}. 
We can also show that $\Psi$ satisfies the equation of motion when it is contracted with any state in the Fock space.  
The energy density of $\Psi$ is the same as that of the perturbative vacuum, 
yet we are not sure whether it is a pure-gauge solution. 
In this case, the formula \eqref{pgf} does not give  a regular $U$, for it 
 contains a negative power of $K$.  
In terms of the winding number\cite{Hata:2011ke}, this is a question 
whether winding numbers around $K=0$ are canceled by those of opposite sign around $K=-\infty$.

\section{Definition of the double-brane solution}
\setcounter{equation}{0}
In this section, we give a definition of the double-brane solution. 
According to the energy formula \eqref{HK_energy1}, 
the following ansatz for the solution is expected to have the energy density of double D-branes:
\begin{equation}\label{new-double-brane-solution2}
\Psi=Kc\frac{K}{1-K}Bc\,.
\end{equation}
The solution \eqref{new-double-brane-solution2} is the 
symmetric counterpart of the following
 under the Hata-Kojita inversion: 
\begin{equation}\label{old-double-brane-solution}
\Psi=\frac{1}{K}c\frac{K^2B}{K-1}c\,.
\end{equation}
This is the familiar ansatz for the double-brane solution \cite{Murata:2011ex}. 
Since it contains the factor $1/K$, the expression \eqref{old-double-brane-solution} itself is clearly singular\,.  
In contrast, the singularity of \eqref{new-double-brane-solution2} is not so clear 
at first glance. 
However, there does exist an unobtrusive singularity,
as essentially explained in \cite{Hata:2012cy}, 
and the energy density of the solution 
is indefinite without suitable regularization.
This seems to be consistent with the discussion by Erler \cite{Erler:2012dz},
for the highest level in the dual $\mathcal L^-$ level expansion of $\Psi$ is zero.

Now, let us present the regularized definition of the solution
in question:  
\begin{equation}\label{regularized_solution}
\begin{split}
\Psi
&= \lim_{\epsilon \to 0}K1_\epsilon  c\frac{KB}{1-K}c\\
&=
\lim_{\epsilon\to 0}
\int_0^\infty dx\,\delta_\epsilon(x) 
\left(\int_0^\infty du\, e^{-u}\frac{\partial}{\partial x}
\frac{\partial}{\partial u}
e^{xK}c e^{uK} Bc\right)\,.\\
\end{split}
\end{equation}
This solution reproduces the energy density of double D-branes. 
The equation of motion 
is satisfied when it is contracted with the solution itself and when it is contracted with any state in the Fock space.   
We also calculate the Ellwood invariant and the boundary state in $\S$ 4.4 and in $\S$ 4.5, respectively. 
We will see that both of them are the same as those for the perturbative vacuum.  

If we change the position where $1_\epsilon$ is inserted, then the properties of the solution
drastically change. 
This means that we can make several distinct solutions with different properties from the ansatz~\eqref{new-double-brane-solution2}.
Note that this situation also occurs when we consider other ansatzes for solutions. 
We will discuss this subject in $\S$ 4.6.

\subsection{Kinetic term}
Let us calculate the normalized kinetic term $\widehat {\mathcal E}_K$ for the solution 
\eqref{regularized_solution},
 defined by
\begin{equation}\notag
\widehat {\mathcal E}_K(\Psi)=\frac{\pi^2}{3}\langle \Psi,\,Q\Psi\rangle\,.
\end{equation}
Note that, if $\Psi$ is a multiple-brane solution, 
the quantity $\widehat {\mathcal E}_K+1$ represents 
the multiplicity of D-branes.
Using the correlation function\footnote
{For an explicit form of this correlation function
\eqref{kinetic-correlation-function}, 
see appendix \ref{notationcorrelation}.}
\begin{equation}\label{kinetic-correlation-function}
C_K(x,y;u,v)\equiv{\rm tr}\left[e^{xK}ce^{uK}Bc\,Q(e^{yK}ce^{vK}Bc)
\right],
\end{equation}
we define the quantity $\mathcal E_K(\eta,\,\epsilon)$ as follows:
\begin{equation}\notag
\mathcal E_K(\epsilon,\,\eta)
=\int_0^\infty du \int_0^\infty dv e^{-u-v} 
\left(\frac{\partial}{\partial x}
\frac{\partial}{\partial y}\frac{\partial}{\partial u}
\frac{\partial}{\partial v}C_K(x,y;u,v)\right)\Bigg |_
{x=\epsilon,\,y=\eta}\,. 
\end{equation}
Thanks to the relation \eqref{trn}, 
the regularized kinetic term can be expressed as\footnote{
To be precise, we need to prove that
$\mathcal E_K(x,\,y)$ satisfies
the conditions 
presented in footnote 8 
before we use the relation~\eqref{trn}.
In appendix~\ref{slcf}, 
we check these conditions.
}
\begin{equation}\label{relation2}
\widehat {\mathcal E}_K=\frac{\pi^2}{3}\frac{1}{2}
\left(\lim_{\epsilon\to 0}\lim_{\eta\to 0}
+\lim_{\eta\to 0}\lim_{\epsilon\to 0}\right)\mathcal E_K(\epsilon,\,\eta)
=\frac{\pi^2}{3}\lim_{\epsilon\to 0}\lim_{\eta\to 0}\mathcal E_K(\epsilon,\,\eta)\,.
\end{equation}
By a straightforward calculation, we find
\begin{equation}\label{regularized kinetic}
\begin{split}
{\mathcal E}_K(\epsilon,\,\eta)
=&\int_0^\infty du \int_0^\infty dv e^{-u-v} 
\left(\frac{\partial}{\partial x}
\frac{\partial}{\partial y}\frac{\partial}{\partial u}
\frac{\partial}{\partial v}C_K(x,y;u,v)\right)
\Bigg |_{x=\epsilon,\,y=\eta} \\
=&\int_0^\infty ds
\frac{4 e^{-s}}{(s+\eta +\epsilon )^8}\\
&\times \Bigg(c(s,\,\epsilon,\,\eta)\cos\left(\frac{2\pi \epsilon}{s+\epsilon+\eta}\right)
+c(s,\,\eta,\,\epsilon)\cos\left(\frac{2\pi \eta}{s+\epsilon+\eta}\right)\\
&-\left((\epsilon+\eta)s^6+2(\epsilon+\eta)^2s^5+(\epsilon+\eta)^3s^4\right)
\cos\left(\frac{2\pi (\epsilon+\eta)}{s+\epsilon+\eta}\right)\\
&+s(s,\,\epsilon,\,\eta)\sin\left(\frac{2\pi \epsilon}{s+\epsilon+\eta}\right)
+s(s,\,\eta,\,\epsilon)\sin\left(\frac{2\pi \eta}{s+\epsilon+\eta}\right)\Bigg)\,.
\end{split}
\end{equation}
Here we changed integration variables from $(u,\ v)$ to 
$(s,\ v)\equiv(u+v,\ v)$. 
The functions $c(s,\,\epsilon,\,\eta)$ 
and $s(s,\,\epsilon,\,\eta)$
 are given by
\begin{equation}\notag
\begin{split}
c(s,\,\epsilon,\,\eta)=
-\epsilon\Big(&s^6+2(\epsilon+\eta)s^5
+(\epsilon^2+5\epsilon \eta-2\eta^2)s^4
+2\eta((4+\pi^2)\epsilon^2-4\eta^2)s^3\\
&+\eta(7\epsilon^3+2(5+2\pi^2)\epsilon^2\eta
-10\epsilon\eta^2-7\eta^3)s^2\\
&+2\eta(\epsilon^4+5\epsilon^3\eta+\pi^2\epsilon^2\eta^2
-5\epsilon\eta^3-\eta^4)s\\
&+3\epsilon(\epsilon-\eta)\eta^2(\epsilon+\eta)^2\Big)\,,
\end{split}
\end{equation}
\begin{equation}\notag
\begin{split}
s(s,\,\epsilon,\,\eta)=
\pi\epsilon^2\Big(
&-s^5-2(\epsilon-\eta)s^4-(\epsilon^2+2\epsilon\eta-12\eta^2)s^3\\
&+2\eta(-2\epsilon^2+2\epsilon\eta+7\eta^2)s^2
-\eta^2(\epsilon^2-6\epsilon\eta-5\eta^2)s\\
&+2\epsilon\eta^3(\epsilon+\eta)
\Big)\,.
\end{split}
\end{equation}
For finite $\epsilon$, we can change the order of 
the $s$-integral and the limit $\eta \rightarrow  0$.
We then obtain that
\begin{equation}\label{flag_kinetic}
\begin{split}
\lim_{\eta\to 0} \mathcal E_K(\epsilon,\,\eta)
&=
-4\pi \epsilon^2\int_0^\infty ds\,\frac{s^3}{(s+\epsilon)^6}e^{-s}
\sin\left(\frac{2 \pi s}{s+\epsilon}\right) \\
&=
-4\pi \epsilon^2
\sum_{j=0}^\infty \frac{(-1)^j}{(2j+1)!}
\int_0^\infty ds\,\frac{s^3}{(s+\epsilon)^6}e^{-s}
\left(\frac{2 \pi s}{s+\epsilon}\right)^{2j+1}\,.
\end{split}
\end{equation}
Using the integration formula below,
\begin{equation}\notag
\int_0^\infty ds\,e^{-s}\frac{s^3}{(s+\epsilon)^6}
\frac{s^{2j+1}}{(s+\epsilon)^{2j+1}}\sim\frac{1}{(2j+5)(2j+6)}
\frac{1}{\epsilon^2}+ O(\epsilon^{-1})\,,
\end{equation}
we find that
\begin{equation}\notag
\begin{split}
\lim_{\epsilon\to 0}\lim_{\eta\to 0}\mathcal E_K(\epsilon,\,\eta)
&= -4\pi \lim_{\epsilon\to 0} \epsilon^2 \sum_{j=0}^{\infty}\frac{(-1)^j}{(2j+1)!}
(2\pi)^{2j+1}\times\frac{1}{(2j+5)(2j+6)}\frac{1}{\epsilon^2}\\
&=-4\pi \times \left(-\frac{3}{4\pi^3}\right)\\
&=\frac{3}{\pi^2}\,.
\end{split}
\end{equation}
Therefore, we conclude that
\begin{equation}\label{pptK}
\widehat {\mathcal E}_K=1\,.
\end{equation}

\subsection{Cubic term}
Let us move on to the cubic term. 
Let us define  the regularized cubic term $\widehat {\mathcal E}_C$ as
\begin{equation}\notag
\widehat {\mathcal E}_C(\Psi)=-\frac{\pi^2}{3}\langle
 \Psi,\,\Psi\ast\Psi\rangle\,.
\end{equation}
Note that the equation of motion 
contracted with $\Psi$ itself is equivalent to the condition $\widehat {\mathcal E}_C(\Psi)=\widehat {\mathcal E}_K(\Psi)$. 
We also define the quantity 
$\mathcal E_C(\epsilon_1,\,\epsilon_2, \,\epsilon_3)$
as follows:
\begin{equation}\notag
\begin{split}
{\mathcal E}_C(\epsilon_1,\,\epsilon_2, \,\epsilon_3)
=&\int_0^\infty du \int_0^\infty dv 
\int_0^\infty dw
e^{-u-v-w} \\
&\left(\frac{\partial}{\partial x}
\frac{\partial}{\partial y}
\frac{\partial}{\partial z}
\frac{\partial}{\partial u}
\frac{\partial}{\partial v}
\frac{\partial}{\partial w}
C_C(x,\,y,\,z;\,u,\,v,\,w)\right)
\Bigg |_{x=\epsilon_1,\,y=\epsilon_2,\,z=\epsilon_3}\,.
\end{split}
\end{equation}
Using \eqref{3rln}, $\widehat {\mathcal E}_C $ is expressed as 
\begin{equation}\label{relation3}
\begin{split}
\widehat {\mathcal E}_C(\Psi)
=&\frac{\pi^2}{3}\frac{1}{3}
\lim_{\epsilon\to0}
\Big(
\mathcal E_C(\epsilon,\,0,\,0)
+
\mathcal E_C(0,\,\epsilon,\,0)
+\mathcal E_C(0,\,0,\,\epsilon)
\Big)
\\
=&\frac{\pi^2}{3}
\lim_{\epsilon\to 0}
\mathcal E_C(\epsilon,\,0,\,0)\,.
\end{split}
\end{equation}
After a straightforward calculation, we find the following expression:
\begin{equation}\notag
\begin{split}
&\mathcal E_C(\epsilon_1,\,\epsilon_2,\,\epsilon_3)\\
=&\int_0^\infty ds
\frac{e^{-s}}{(s+\epsilon_1+\epsilon_2+\epsilon_3 )^9}\\
&\Bigg(
c_1(s,\,\epsilon_1,\,\epsilon_2,\,\epsilon_3)
\cos\left(\frac{2\pi \epsilon_1}
{s+\epsilon_1+\epsilon_2+\epsilon_3}\right)
+
c_1(s,\,\epsilon_2,\,\epsilon_3,\,\epsilon_1)
\cos\left(\frac{2\pi \epsilon_2}
{s+\epsilon_1+\epsilon_2+\epsilon_3}\right)\\
&+
c_1(s,\,\epsilon_3,\,\epsilon_1,\,\epsilon_2)
\cos\left(\frac{2\pi \epsilon_3}
{s+\epsilon_1+\epsilon_2+\epsilon_3}\right)
+
c_2(s,\,\epsilon_1,\,\epsilon_2,\,\epsilon_3)
\cos\left(\frac{2\pi (\epsilon_1+\epsilon_2)}
{s+\epsilon_1+\epsilon_2+\epsilon_3}\right)\\
&+
c_2(s,\,\epsilon_2,\,\epsilon_3,\,\epsilon_1)
\cos\left(\frac{2\pi (\epsilon_2+\epsilon_3)}
{s+\epsilon_1+\epsilon_2+\epsilon_3}\right)
+
c_2(s,\,\epsilon_3,\,\epsilon_1,\,\epsilon_2)
\cos\left(\frac{2\pi (\epsilon_3+\epsilon_1)}
{s+\epsilon_1+\epsilon_2+\epsilon_3}\right)\\
&+
c_3(s,\,\epsilon_1,\,\epsilon_2,\,\epsilon_3)
\cos\left(\frac{2\pi (\epsilon_1+\epsilon_2+\epsilon_3)}
{s+\epsilon_1+\epsilon_2+\epsilon_3}\right)
+
s_1(s,\,\epsilon_1,\,\epsilon_2,\,\epsilon_3)
\sin\left(\frac{2\pi \epsilon_1}
{s+\epsilon_1+\epsilon_2+\epsilon_3}\right)\\
&+
s_1(s,\,\epsilon_2,\,\epsilon_3,\,\epsilon_1)
\sin\left(\frac{2\pi \epsilon_2}
{s+\epsilon_1+\epsilon_2+\epsilon_3}\right)
+
s_1(s,\,\epsilon_3,\,\epsilon_1,\,\epsilon_2)
\sin\left(\frac{2\pi \epsilon_3}
{s+\epsilon_1+\epsilon_2+\epsilon_3}\right)\\
&+
s_2(s,\,\epsilon_1,\,\epsilon_2,\,\epsilon_3)
\sin\left(\frac{2\pi (\epsilon_1+\epsilon_2)}
{s+\epsilon_1+\epsilon_2+\epsilon_3}\right)
+
s_2(s,\,\epsilon_2,\,\epsilon_3,\,\epsilon_1)
\sin\left(\frac{2\pi (\epsilon_2+\epsilon_3)}
{s+\epsilon_1+\epsilon_2+\epsilon_3}\right)\\
&+
s_2(s,\,\epsilon_3,\,\epsilon_1,\,\epsilon_2)
\sin\left(\frac{2\pi (\epsilon_3+\epsilon_1)}
{s+\epsilon_1+\epsilon_2+\epsilon_3}\right)
+
s_3(s,\,\epsilon_1,\,\epsilon_2,\,\epsilon_3)
\sin\left(\frac{2\pi (\epsilon_1+\epsilon_2+\epsilon_3)}
{s+\epsilon_1+\epsilon_2+\epsilon_3}\right)\Bigg)\,,
\end{split}
\end{equation}
where
\begin{equation}\notag
\begin{split}
&c_1(s,\,x,\,y,\,z)\\
=
&4 x^2 \big(
3 s^6+6 s^5 x+9 s^5 y+9 s^5 z+3 s^4 x^2
+17 s^4 x y+17 s^4 x z+6 s^4
   y^2+12 s^4 y z+6 s^4 z^2\\
&+2 \pi ^2 s^3 x^2 y+15 s^3 x^2 y
+2 \pi ^2 s^3 x^2 z+15 s^3
   x^2 z+12 s^3 x y^2+24 s^3 x y z+12 s^3 x z^2-6 s^3 y^3\\
&-18 s^3 y^2 z-18 s^3 y z^2-6
   s^3 z^3+9 s^2 x^3 y+9 s^2 x^3 z+4 \pi ^2 s^2 x^2 y^2+18 s^2 x^2 y^2+8 \pi ^2 s^2
   x^2 y z\\
&+36 s^2 x^2 y z+4 \pi ^2 s^2 x^2 z^2+18 s^2 x^2 z^2-6 s^2 x y^3-18 s^2 x
   y^2 z-18 s^2 x y z^2-6 s^2 x z^3-9 s^2 y^4\\
&-36 s^2 y^3 z-54 s^2 y^2 z^2
-36 s^2 y
   z^3-9 s^2 z^4+2 s x^4 y+2 s x^4 z+12 s x^3 y^2+24 s x^3 y z+12 s x^3 z^2\\
&+2 \pi ^2
   s x^2 y^3+3 s x^2 y^3+6 \pi ^2 s x^2 y^2 z+9 s x^2 y^2 z+6 \pi ^2 s x^2 y z^2+9 s
   x^2 y z^2+2 \pi ^2 s x^2 z^3+3 s x^2 z^3\\
&-10 s x y^4-40 s x y^3 z-60 s x y^2 z^2-40
   s x y z^3-10 s x z^4-3 s y^5-15 s y^4 z-30 s y^3 z^2\\
&-30 s y^2 z^3-15 s y z^4-3 s
   z^5+3 x^4 y^2+6 x^4 y z+3 x^4 z^2+3 x^3 y^3+9 x^3 y^2 z+9 x^3 y z^2+3 x^3 z^3\\
&-3
   x^2 y^4-12 x^2 y^3 z-18 x^2 y^2 z^2-12 x^2 y z^3-3 x^2 z^4-3 x y^5-15 x y^4 z-30 x
   y^3 z^2-30 x y^2 z^3\\
&-15 x y z^4-3 x z^5\big)\,,
\end{split}
\end{equation}

\begin{equation}\notag
\begin{split}
&c_2(s,\,x,\,y,\,z)\\
=&4 (s+z) (x+y) \big(
s^6+2 s^5 z-3 s^4 x^2-6 s^4 x y-3 s^4 y^2-2 s^4 z^2-2 s^3 x^3-6
   s^3 x^2 y+2 \pi ^2 s^3 x^2 z\\
&-6 s^3 x y^2+4 \pi ^2 s^3 x y z-3 s^3 x z^2-2 s^3
   y^3+2 \pi ^2 s^3 y^2 z-3 s^3 y z^2-8 s^3 z^3+2 s^2 x^3 z+6 s^2 x^2 y z\\
&+4 \pi ^2
   s^2 x^2 z^2+6 s^2 x^2 z^2+6 s^2 x y^2 z+8 \pi ^2 s^2 x y z^2+12 s^2 x y z^2-9 s^2
   x z^3+2 s^2 y^3 z+4 \pi ^2 s^2 y^2 z^2\\
&+6 s^2 y^2 z^2-9 s^2 y z^3-7 s^2 z^4+7 s x^3
   z^2+21 s x^2 y z^2+2 \pi ^2 s x^2 z^3+21 s x y^2 z^2+4 \pi ^2 s x y z^3\\
&-9 s x
   z^4+7 s y^3 z^2+2 \pi ^2 s y^2 z^3-9 s y z^4-2 s z^5+3 x^4 z^2+12 x^3 y z^2+3 x^3
   z^3+18 x^2 y^2 z^2\\
&+9 x^2 y z^3-3 x^2 z^4+12 x y^3 z^2+9 x y^2 z^3-6 x y z^4-3 x
   z^5+3 y^4 z^2+3 y^3 z^3-3 y^2 z^4-3 y z^5
\big)\,,
\end{split}
\end{equation}

\begin{equation}\notag
\begin{split}
c_3(s,\,x,\,y,\,z)=-4(s+x+y+z)^2 s^4 (2s-x-y-z)(x+y+z)\,,
\end{split}
\end{equation}

\begin{equation}\notag
\begin{split}
&s_1(s,\,x,\,y,\,z)\\
=&
-4 \pi  x^3 (s+y+z) (-s-x-y-z) \big(2 s^3+s^2 x-4 s^2 y-4 s^2 z+3 s x y+3 s x z-6 s
   y^2\\
&-12 s y z-6 s z^2-2 x y^2-4 x y z-2 x z^2\big)\,,
\end{split}
\end{equation}

\begin{equation}\notag
\begin{split}
&s_2(s,\,x,\,y,\,z)\\
=&4 \pi  e^{-s} (s+y)^2 (x+z)^2 (-s-x-y-z) \left(s^3-4 s^2 y+4 s x y-5 s y^2+4 s y z-2
   x y^2-2 y^2 z\right)\,,
\end{split}
\end{equation}

\begin{equation}\notag
s_3(s,\,x,\,y,\,z)=4 \pi  s^5 (x+y+z)^2 (s+x+y+z)\,.
\end{equation}
As far as we keep $\epsilon_1$ finite, we  
can take the limits $\epsilon_2 \to 0$ and
$\epsilon_3 \to 0$ before we perform 
the $s$ integral:
\begin{equation}\notag
\lim_{\epsilon_2 \to 0}\lim_{\epsilon_3 \to 0}
\mathcal E_C(\epsilon_1,\,\epsilon_2,\,\epsilon_3)=
-4\pi\epsilon_1^2\int_0^\infty ds\,
\frac{s^3}{(s+\epsilon_1)^6}e^{-s}
\sin\left(\frac{2\pi\epsilon_1}{s+\epsilon_1}\right)\,.
\end{equation}
This integral is the same as that appearing in \eqref{flag_kinetic}. 
Therefore, we obtain
\begin{equation}\label{pptC}
\widehat {\mathcal E}_C=1.
\end{equation}
From \eqref{pptK} and \eqref{pptC}, we conclude that the energy density of the solution \eqref{regularized_solution} is that of the double D-brane. 

\subsection{Equation of motion}

So far we have confirmed that the solution reproduces the energy density for 
double D-branes. 
From \eqref{pptK} and \eqref{pptC}, we can also conclude that 
the equation of motion 
is satisfied 
when it is contracted with the solution itself,
\begin{equation}\notag
{\rm tr}[\,\Psi\, Q \Psi\,]=
-{\rm tr}[\,\Psi\, \Psi\, \Psi\,]
\quad \left(=\frac{3}{\pi^2}\right)\,.
\end{equation}
Now, let us investigate the equation of motion contracted with 
states in the Fock space.
It is apparently satisfied,
for states in the Fock space always can be 
written as a wedge state of width one 
with local operator insertions. 

Let $\phi$ be a state in the Fock space.
Each term of the equation of motion can be
 written as follows:  
\begin{equation}\notag
\begin{split}
{\rm tr}[\,\Psi\Psi\,\phi\,]
=&
{\rm tr}\left[\,Kc\frac{KB}{1-K}cKc\frac{KB}{1-K}c\,\phi\,\right]\\
=&
\int_0^\infty du\, \int_0^\infty dv\,
e^{-u-v}\frac{\partial}{\partial x}\frac{\partial}{\partial u}
\frac{\partial}{\partial y}\frac{\partial}{\partial v}
{\rm tr}[\,e^{xK}ce^{uK}Bce^{yK}ce^{vK}Bc\,\phi\,]\\
=&
\int_0^\infty du\, \int_0^\infty dv\,
e^{-u-v}\frac{\partial}{\partial x}\frac{\partial}{\partial u}
\frac{\partial}{\partial y}\frac{\partial}{\partial v}\\
&\times\left({\rm tr}[\,e^{xK}ce^{uK}e^{yK}ce^{vK}Bc\,\phi\,]
-{\rm tr}[\,e^{xK}ce^{uK}ce^{yK}e^{vK}Bc\,\phi\,]
\right )\\
=&
\int_0^\infty du\, \int_0^\infty dv\,
e^{-u-v}
\left(
C_\phi^{(1,\,2,\,1)}(0,u,v)-C_\phi^{(1,\,1,\,2)}(0,u,v)
\right)
\,,
\end{split}
\end{equation}
\begin{equation}\notag
\begin{split}
{\rm tr}[\,(Q\Psi)\,\phi\,]
=&
{\rm tr}\left[\,\left(Q Kc\frac{KB}{1-K}c\right)\,\phi\,\right]\\
=&
\int_0^\infty du\,
e^{-u}\frac{\partial}{\partial x}\frac{\partial}{\partial u}
{\rm tr}\left[\,\left(Qe^{xK}ce^{uK}Bc\right)\,\phi\,\right]\\
=&
\int_0^\infty du\,
e^{-u}\frac{\partial}{\partial x}\frac{\partial}{\partial u}
\left({\rm tr}[\,e^{xK}cKce^{uK}Bc\,\phi\,]
-{\rm tr}[\,e^{xK}ce^{uK}cKBc\,\phi\,]
\right )\\
=&
\int_0^\infty du\,
e^{-u}
\left(C_\phi^{(1,\,1,\,1)}(0,\,0,\,u)-
C_\phi^{(1,\,1,\,1)}(0,\,u,\,0)
\right )\,,
\end{split}
\end{equation}
where we defined 
\begin{equation}\notag
C_\phi(x,\,y,\,z)\equiv{\rm tr}[\,e^{xK}ce^{yK}ce^{zK}Bc\,\phi\,]\,,
\end{equation}
and
\begin{equation}\notag
C_\phi^{(i,\,j,\,k)}(x_0,\,y_0,\,z_0)\equiv
\frac{\partial^i}{\partial x^i}
\frac{\partial^j}{\partial y^j}
\frac{\partial^k}{\partial z^k}
C_\phi(x,\,y,\,z)\Big|_{x=x_0,\,y=y_0,\,z=z_0}\,.
\end{equation}
Above expressions are valid as far as $C_\phi(x,\,y,\,z)$ is 
analytic around $(x,\,y,\,z)=(0,\,u,\,v)$, $(0,\,0,\,u)$ 
and $(0,\,u,\,0)$.
Since $C_\phi(x,\,y,\,z)$ is a correlation function of 
three local operator insertions with a line integral of $b$-ghost,
$C_\phi(x,\,y,\,z)$ is regular for $0\le x+y+z<\infty$.
Using integration by parts,
\begin{equation}\notag
\begin{split}
\int_0^\infty du\,e^{-u}C_\phi^{(1,\,2,\,1)}(0,\,u,\,v)
=&-C_\phi^{(1,\,1,\,1)}(0,\,0,\,v)
+\int_0^\infty du\,e^{-u}C_\phi^{(1,\,1,\,1)}(0,\,u,\,v)\,,
\end{split}
\end{equation}
and
\begin{equation}\notag
\begin{split}
\int_0^\infty dv\,e^{-v}C_\phi^{(1,\,1,\,2)}(0,\,u,\,v)
=&-C_\phi^{(1,\,1,\,1)}(0,\,u,\,0)
+\int_0^\infty dv\,e^{-v}C_\phi^{(1,\,1,\,1)}(0,\,u,\,v)\,,
\end{split}
\end{equation}
we conclude that $\Psi$ satisfies the equation of motion contracted with any state $\phi$ in the Fock space:
\begin{equation}\notag
{\rm tr}[\,(Q\Psi)\,\phi\,]+{\rm tr}[\,\Psi\Psi\,\phi\,]=0.
\end{equation}

\subsection{The Ellwood invariant}
\label{el}
In \cite{Ellwood:2008jh}, Ellwood conjectured that there exists a relation between the gauge-invariant 
observables of open string field theory which were discovered in \cite{HI,GRSZ}, 
and 
the closed string tadpole on a disk.\footnote{In\cite{Baba:2012cs,HKE}, this conjecture was 
investigated in detail for the special case where $\phi_\text{closed}$ is a graviton. 
In particular, some correction to this relation was proposed in \cite{HKE}.  
} 
In this paper, we call these gauge-invariant observables the Ellwood invariant. 

The Ellwood invariant for a classical solution $\Psi$ is defined by 
\begin{equation}\notag
\mathcal W(\Psi,\, \phi_{\text{closed}})=
\Big\langle\, \phi_{\text{closed}}(i)\ f_I\circ \Psi(0)\,\Big
\rangle_{\text{UHP}}\,.
\end{equation}
Here, $\phi_{\text{closed}}$ is a closed string vertex operator of weight (1,1) and ghost number 2;
$\Psi(0)$ is the operator corresponding to the classical solution $\Psi$, and $f_I\circ \Psi(0)$ is the conformal transformation of $\Psi(0)$ under the map associated with the identity state, 
\begin{equation}\notag
f_I(\xi)\equiv\frac{2\xi}{1-\xi^2}\,.
\end{equation}
Ellwood conjectured that $\mathcal{W} (\Psi,\,\phi_{\text{closed}})$ is 
equivalent to the difference of two tadpole diagrams, 
\[\mathcal{W} (\Psi,\,\phi_{\text{closed}})= \mathcal A_\Psi(\phi_{\text{closed}})
-\mathcal A_0(\phi_{\text{closed}})\,.\]
Here $\mathcal A_0(\phi_{\text{closed}})$ denotes the closed string tadpole 
on a disk 
with the original boundary condition, and $\mathcal A_\Psi(\phi_{\text{closed}})$ denotes the closed string tadpole 
with the boundary condition  corresponding to the classical solution $\Psi$. 

In \cite{Murata:2011ex,Murata:2011ep}, 
Murata and Schnabl calculated the Ellwood invariant for the Okawa-type solution~\eqref{Okawa-solution}. 
Simply applying their formula to the solution \eqref{regularized_solution}, we find that the Ellwood invariant for the solution %
 is zero,  
\begin{equation}\notag
\mathcal{W} (\Psi,\,\phi_{\text{closed}})=-\lim_{\epsilon\to 0}\left(
\lim_{z\to 0}\frac{d F_\epsilon(z)^2}{dz}H_\epsilon(z)\right)
\mathcal A_0(\phi_{\text{closed}})=0\,,
\end{equation}
where
\begin{equation}\notag
F_\epsilon(K)^2=\int_0^\infty dx\, \delta_\epsilon(x)\,e^{x K}K\,,
\qquad H_\epsilon(K)=\frac{K}{1-K}\,.
\end{equation}
This means that the Ellwood invariant of the solution $\Psi$ is that of the perturbative vacuum.

\subsection{Boundary states}
\label{bo}
In \cite{KOZ}, Kiermaier, Okawa and Zwiebach constructed a closed string state 
$|\,B_\ast(\Psi)\,\rangle$ 
from classical solutions $\Psi$ of open string field theory. 
The closed string state is invariant under the gauge transformations of $\Psi$. 
For several known solutions, $|\,B_\ast(\Psi)\,\rangle$ corresponds to 
the boundary state of the vacuum  
which the classical solution $\Psi$ represents. 
We here simply refer to the closed string state $|\,B_\ast(\Psi)\,\rangle$ 
as the boundary state. 

In \cite{Takahashi:2011wk,Masuda:2012kt}, the boundary states for 
different classical solutions 
in the $KBc$ subalgebra are calculated. 
Let $|\,B\,\rangle$ denote the boundary state for the perturbative vacuum. 
The boundary state for the Okawa-type solution \eqref{Okawa-solution} is given by  
\[|\,B_*(\Psi_F)\,\rangle=\frac{e^{(x+1)s}-e^{ys}}{e^s-1}|\,B\,\rangle\,,\]
where\footnote{These expressions for $x$ and $y$ are valid only for the non-real solution \eqref{Okawa-solution}. For more general expression, see \cite{Masuda:2012kt}. }
\[x=
\frac{z}{1-F(z)^2}
\left(\frac{1}{2}F(z)^2+2F'(z)F(z)\right)\bigg|_{z= 0}\,,
\]
\[
y=\frac{z}{1-F(z)^2}
\left(\frac{1}{2}F(z)^2\right)
\bigg|_{z= 0}\,.\]
From this formula, we find that 
the boundary state $|\,B_\ast(\Psi)\,\rangle$ 
for \eqref{regularized_solution} is that of  
the perturbative vacuum,
\begin{equation}\notag
|\,B_\ast(\Psi)\,\rangle=|\,B\,\rangle\,.\end{equation}

\subsection{Remarks on the ambiguity of classical solutions}
Now, let us 
slightly modify the definition of the solution 
\eqref{regularized_solution}. 
We consider the following solution: 
\begin{equation}\label{perturbative-ordering}
\Psi=
\lim_{\epsilon\to 0}
 Kc1_\epsilon\frac{KB}{1-K}c\,.
\end{equation}
It is straightforward to calculate the energy or the Ellwood invariant 
of $\Psi$. 
We summarize properties of this solution as follows:
\begin{itemize}
\item The energy density of this solution is zero. 
\item The equation of motion is satisfied when it is contracted 
to the solution itself and when it is contracted with states in the Fock space.
\item The Ellwood invariant is for the perturbative vacuum.
\end{itemize}
At least naively, both \eqref{perturbative-ordering}
 and \eqref{regularized_solution} can be 
considered as regularizations of  \eqref{new-double-brane-solution2}.
To define the solution without ambiguity,
we need to regularize the solution and determine 
the order of limits.
Note that these two solutions, \eqref{perturbative-ordering}
 and \eqref{regularized_solution}, possess the same components 
at every level.


This kind of ambiguity is not limited to the ansatz \eqref{new-double-brane-solution2}. 
Take the identity-based solution $-(1+K)c$ for example.  
As stated in $\S$ 3.2, it can be regularized as \eqref{id}. 
On the other hand, as described in Zeze\cite{Zeze}, we can also regularize it using the one parameter family of tachyon vacuum solutions 
that interpolates $(1-K)c$ and the simple tachyon-vacuum solution as follows:
\begin{equation}\label{Zeze}
\Psi=-\lim_{\epsilon\to 0}\frac{1-(\epsilon-1)K}{1-\epsilon K}c(1-\epsilon K)Bc\,.
\end{equation}
Two regularized solutions, \eqref{id} and \eqref{Zeze}, are different in physical properties. 
The energy density of \eqref{id} is zero, while that of \eqref{Zeze} is 
$-1/(2\pi^2g_o^2)$\,. 
It is hoped to gain a deeper understanding of different regularization methods and 
be able to predict the properties of the regularized solutions without calculating the physical quantities.

Let us here state one more question about the solution \eqref{regularized_solution}. 
The expression \eqref{new-double-brane-solution2} can formally be written as a pure-gauge form as follows: 
\begin{equation}\notag
\Psi=UQU^{-1}\,,
\end{equation}
where
\begin{equation}\notag
U=1-KBc\,, \qquad
U^{-1}=1+\frac{K}{1-K}Bc\,.
\end{equation}
If we define the solution as 
\eqref{regularized_solution}, 
we expect that the solution is not true pure gauge. 
So, the gauge parameter $U$ or $U^{-1}$ must be singular in some sense. In particular, they must be  
disconnected to $1$. 
We need to understand in what sense it is singular and characterize 
the singularity.

\section{Summary}
\setcounter{equation}{0}
We presented the double-brane solution \eqref{sln1} based on the ansatz of Hata and
Kojita. 
The solution possesses finite energy density, 
which corresponds to the energy density of double D-branes.  
We also checked that the solution satisfies the equation of motion 
when it is contracted with the solution itself, and when it is contracted with any state of the Fock space. 
However, the Ellwood invariant and the boundary state
are those for the perturbative vacuum. 
These inharmonious results make the physical interpretation of the solution difficult. 
Further research will be needed before the solution is fully 
accepted. 
In particular, we need to clarify the relation 
of our results and the discussion by Baba and Ishibashi 
\cite{Baba:2012cs},
where the authors proved the correspondence between the energy density and 
the Ellwood invariant of classical solutions in part.
It is also important to calculate the boundary state using 
the newly-proposed method by Kudrna, Maccaferri and Schnabl \cite{Kudrna:2012re}.

\ 

\noindent
{\bf \large Acknowledgments}

\noindent
I would like to thank Yuji~Okawa for valuable discussion 
and for detailed reading of the manuscript. 
I am indebted to Theodore Erler for valuable discussion and especially for  
 important comments on my work, 
 which helped me to correct some critical errors. 
I also would like to thank Toshiko~Kojita, 
Toshifumi~Noumi
and
Daisuke~Takahashi for 
valuable and useful discussion. 
I also would like to thank Masaki Murata and Martin Schnabl
 for valuable discussion on the solution \eqref{old-double-brane-solution} and the anomaly of the equation of motion. 
I also would like to thank 
Hiroyuki Hata, 
Yuki Iimori, 
Mitsuhiro Kato, 
Shota Komatsu, 
Koichi Murakami and Shingo~Torii  for valuable discussion. 
I also thank Akiko Maruyama and Takayuki Yanagi 
for 
stimulating conversation. 
I acknowledge that  the regularization method in $\S$ 3 and $\S$ B.1 is inspired by 
the formula for numerical integration of Iri, Moriguti and Takasawa\cite{IMT}. 
Finally, I would like to thank the referee for offering constructive suggestions leading to improvements of the manuscript.

\bigskip

\appendix
\section{Proof of \eqref{conv1} and \eqref{conv2}}
\setcounter{equation}{0}
\label{epsilondelta}
Let us prove \eqref{conv1} under the assumption presented in footnote 8.  
Using the $\epsilon$-$\delta$ definition of limit,  the statement $f_0(x,\, y)\equiv \lim_{r\to 0}f(xr,\, yr)$ is expressed as 
\[^\forall \varepsilon>0,\ ^\exists \delta_1(\varepsilon)>0 \text{ such that }0<r<\delta_1(\varepsilon)\to
\left|f(x r,\ y r)-f_0(x,\ y)\right|<\varepsilon\,.\]
Since $f_0(x,\,y)$ is continuous at $(x,\ y)=(0,\ 1)$, it follows that 
\[
^\forall \varepsilon>0,\ ^\exists \delta_2(\varepsilon)>0 \text{ such that }0<r<\delta_2(\varepsilon)\to
\left|f_0(r,\ 1)-f_0(0,\ 1)\right|<\varepsilon\,.\]
From the conditions 1, 2 and 3 in \S {3.1}, we also have
\[
^\forall \varepsilon>0,\ ^\exists \delta_3(\varepsilon)>0 \text{ such that }0<r<\delta_3(\varepsilon)\to
\left|\frac{\lambda_r(h_1)}{\lambda_r(h_2)}\right|<\varepsilon\,,\]
and
\[
^\forall \varepsilon>0,\ ^\exists \delta_4(\varepsilon)>0 \text{ such that }0<r<\delta_4(\varepsilon)\to
\left|{\lambda_r(h_2)}\right|<\varepsilon\,.\]
Now, setting    
$\delta_5(\epsilon)\equiv \text{min}\left\{\delta_4(\delta_1(\varepsilon/2)),\ \delta_3(\delta_2(\varepsilon/2))\right\}$, 
it follows that 
\begin{equation}\notag
\begin{split}
0<r<\delta_5(\epsilon)
\to\quad
&\left|f(\lambda_r(h_1),\,\lambda_r(h_2))-f_0(0,\,1)\right|\\
<&
\left|
f\left(
\frac{\lambda_r(h_1)}{\lambda_r(h_2)}\lambda_r(h_2),\,\lambda_r(h_2)
\right)-f_0\left(\frac{\lambda_r(h_1)}{\lambda_r(h_2)},\,1\right)\right|
+
\left|f_0\left(\frac{\lambda_r(h_1)}{\lambda_r(h_2)},\,1\right)-f_0(0,\,1)\right|\\
<&
\epsilon\,.
\end{split}
\end{equation}
Thus, noting $\lim_{a\to 0}f(0,\,a)=f_0(0,\,1)$, we obtain \eqref{conv1}. 
In like manner, we can prove \eqref{conv2} and (3.4). 

\section{Some limits of correlation functions}
\setcounter{equation}{0}
\label{slcf}
In this appendix, we explicitly calculate
$
\lim_{\epsilon\to 0}\mathcal E_K(a \epsilon,\,b \epsilon)
$ 
and
$
\lim_{\epsilon\to 0}\mathcal E_C(a \epsilon,\,b \epsilon,\,c \epsilon)\,.
$ 
We start from the expression \eqref{regularized kinetic}.
We take up the first term and consider the following limit:
for $0\le k \le 6$,
\begin{equation}
\begin{split}
&\lim_{\epsilon\to 0}
\int_0^\infty ds\, \frac{s^k (a\epsilon)^{7-k}}
{(s+a\epsilon+b \epsilon)^8}
e^{-s}\cos\left (\frac{2\pi a\epsilon}
{s+a\epsilon+b\epsilon} \right)\\
=
&\lim_{\epsilon\to 0}
\int_0^\infty ds\, \frac{s^k }
{(s+\alpha)^8}
e^{-\epsilon s}\cos\left (\frac{2\pi }
{s+\alpha} \right)\\
=
&\lim_{\epsilon\to 0}
\sum_{l=0}^\infty \frac{(-)^l}{(2l)!}
\int_0^\infty ds\, \frac{s^k }
{(s+\alpha)^8}
e^{-\epsilon s}\left (\frac{2\pi }
{s+\alpha} \right)^{2l}\,,
\end{split}
\end{equation}
where we put $\alpha=1+b/a$. 
For $k=0$, the integral in this 
expression can be written as
\begin{equation}\label{lll}
\int_0^\infty ds\,\frac{1}{(s+\alpha)^8}e^{-\epsilon s}
\left(
\frac{1}{s+\alpha}
\right)^{2l}
=e^{\alpha \epsilon} \epsilon^{7+2l}
\Gamma(-7-2l,\,\alpha\epsilon)\,,
\end{equation}
where $\Gamma(z,\,\epsilon)$ denotes
 the incomplete gamma function
defined by
\begin{equation}\notag
\Gamma(z,\,\epsilon)\equiv \int_\epsilon^\infty e^{-t}t^{z-1}dt.
\end{equation}
We now differentiate \eqref{lll} with respet to $\epsilon$. 
Using the relations,
\begin{equation}\notag
\lim_{\epsilon\to 0}
\epsilon^k \Gamma(-k,\epsilon) = \frac{1}{k}\quad (k  \in \mathbb{N})\,,
\end{equation}
and
\begin{equation}\notag
\begin{split}
\frac{d^m}{d\epsilon^m}\left[\epsilon^k \Gamma(-k,\epsilon) \right]
&=(-1)^m\epsilon^{k-m}\Gamma(-k+m,\,\epsilon)\,,
\end{split}
\end{equation}
we obtain that
\begin{equation}
\begin{split}
&\lim_{\epsilon\to 0}
\int_0^\infty ds\, \frac{s^k (a\epsilon)^{7-k}}
{(s+a\epsilon+b \epsilon)^8}
e^{-s}\cos\left (\frac{2\pi a\epsilon}
{s+a\epsilon+b\epsilon} \right)\\
=&(-1)^k\sum_{l=0}^\infty \frac{(2\pi i)^{2l}}{(2l)!}
\alpha^{k-2l-7}\sum_{j=0}^k \left( \begin{array}{c}
k\\
j
\end{array}\right)\frac{(-1)^j}{7+2l-j}\\
=&k!\sum_{l=0}^\infty \frac{(2\pi i)^{2l}}{(2l)!}
\frac{(2l+6-k)!}{(2l+7)!}\alpha^{k-2l-7}\,.
\end{split}
\end{equation}
Note that this series can be expressed in terms of  
trigonometric functions.

Similarly, we can derive the following expressions: for $0\le k\le 6$,
\begin{equation}
\begin{split}
&\lim_{\epsilon\to 0}\int_0^\infty ds\,
\frac{s^k(a\epsilon)^{7-k}}{(s+a\epsilon + b\epsilon )^8}
e^{- s}\sin\left(\frac{2\pi a\epsilon}
{s+a\epsilon+b\epsilon}\right)
=k!\sum_{l=0}^\infty \frac{2\pi(2\pi i)^{2l}}{(2l+1)!}
\frac{(2l+7-k)!}{(2l+8)!}\alpha^{k-2l-8}\,,
\end{split}
\end{equation}

\begin{equation}
\begin{split}
&\lim_{\epsilon\to 0}\int_0^\infty ds\,
\frac{s^k(a\epsilon+b\epsilon)^{7-k}}{(s+a\epsilon + b\epsilon )^8}
e^{- s}\cos\left(\frac{2\pi(a\epsilon+b\epsilon)}
{s+a\epsilon+b\epsilon}\right)
=\sum_{l=0}^\infty \frac{(2\pi i)^l}{(2l)!}
\frac{(2l+k)!(6-k)!}{(2l+7)!}\,,
\end{split}
\end{equation}

\begin{equation}
\begin{split}
&\lim_{\epsilon\to 0}\int_0^\infty ds\,
\frac{s^k(a\epsilon+b\epsilon)^{7-k}}{(s+a\epsilon + b\epsilon )^8}
e^{- s}\sin\left(\frac{2\pi(a\epsilon+b\epsilon)}
{s+a\epsilon+b\epsilon}\right)
=\sum_{l=0}^\infty \frac{(2\pi i)^{2l+1}i}{(2l+1)!}
\frac{(2l+k+1)!(6-k)!}{(2l+8)!}\,.
\end{split}
\end{equation}
Using above formulae, we obtain the following expressions:
\begin{equation}\label{last_expressions1}
\begin{split}
\lim_{\epsilon\to 0}\mathcal E_K(a \epsilon,\,b \epsilon)
=&\frac{2}{\pi^2}+\frac{(a+b)^2+2\pi^2 ab}{\pi^2(a+b)^2}
\cos\left(\frac{2a\pi}{a+b}\right)
+\frac{a-b}{\pi(a+b)}\sin\left(\frac{2a\pi}{a+b}\right)\,,
\end{split}
\end{equation}

\begin{equation}\label{last_expressions2}
\begin{split}
\lim_{\epsilon\to 0}\mathcal E_C(a \epsilon,\,b \epsilon,\,c\epsilon)
=&-\frac{3}{\pi^2}
-\frac{(2a-b-c)(a+b+c)^2+2\pi^2 a^2(b+c)}{\pi^2(a+b+c)^3}
 \cos\left(\frac{2\pi a}{a+b+c}\right)\\
 &
-\frac{(2b-c-a)(a+b+c)^2+2\pi^2 b^2(c+a)}{\pi^2(a+b+c)^3}
 \cos\left(\frac{2\pi b}{a+b+c}\right)\\
 &
-\frac{(2c-a-b)(a+b+c)^2+2\pi^2 c^2(a+b)}{\pi^2(a+b+c)^3}
 \cos\left(\frac{2\pi c}{a+b+c}\right)\\
 &
+\frac{3(a+b+c)^2-2\pi^2 a (a-2b-2c)}{2\pi^3(a+b+c)^2}
 \sin\left(\frac{2\pi a}{a+b+c}\right)\\
  &
+\frac{3(a+b+c)^2-2\pi^2 b (b-2c-2a)}{2\pi^3(a+b+c)^2}
 \sin\left(\frac{2\pi b}{a+b+c}\right)\\
  &
+\frac{3(a+b+c)^2-2\pi^2 c (c-2a-2b)}{2\pi^3(a+b+c)^2}
 \sin\left(\frac{2\pi c}{a+b+c}\right)\,.
\end{split}
\end{equation}
From these expressions, we see that $\mathcal E_K(x,\,y)$ satisfies the condition presented in 
footnote~8, and 
$\mathcal E_C(x,\,y,\,z)$ satisfies the condition presented in footnote~9, respectively.

\section{On the ansatz \eqref{old-double-brane-solution} for the double-brane solution}
\setcounter{equation}{0}
Following Murata and Schnabl\cite{Murata:2011ex}, 
several studies have been made to 
construct the multiple-brane solutions  based on~\eqref{Murata-Schnabl energy}\cite{Murata:2011ep,Hata:2011ke}.  
The point here is that the expression 
 \eqref{old-double-brane-solution} contains a factor 1/$K$, and we need to regularize it. 
In this appendix, we summarize our attempt to construct the double-brane solution based on the ansatz~\eqref{old-double-brane-solution}. 
We show that the regularized solution satisfies the equation of motion when it is contracted with the solution itself. 
We also show that the equation of motion is broken when it is contracted with some 
states in the Fock space. 
These results are similar as those of \cite{Murata:2011ep,Hata:2011ke}, 
where the solution is regularized using the $\epsilon$-regularization.

\subsection{Regularization}
Consider a string field $\varphi(\Lambda)$  
with a large cutoff parameter~$\Lambda$. 
We define a regularized string field $\varphi_R$ as follows:
\begin{equation}\label{regularized_varphi}
\varphi_R\equiv \lim_{\Lambda\to\infty}
\varphi_R (\Lambda)
\quad\ 
{\rm with}
\quad\ 
\varphi_R (\Lambda)\equiv \int_0^1 ds
\, \varphi(\lambda(\Lambda;\,s))\,,
\end{equation}
where
\begin{equation}
\lambda(\Lambda;\,s)\equiv(\Lambda +1)^s-1,\qquad 0\le s\le 1\,.
\end{equation}
The following property is important for our discussion:
\begin{equation}\label{imp}
\lim_{\Lambda\to\infty}
\frac{\lambda(\Lambda;\,s_1)}{\lambda(\Lambda;\,s_2)}
=\begin{cases}
\infty\qquad& s_1>s_2\,,\\
0      \qquad& s_2>s_1\,.
\end{cases}
\end{equation}

We now would like to prove an identity which is  similar to \eqref{trn}. 
Let  $\tilde f( \varphi(\Lambda_1),\,\varphi(\Lambda_2))$ 
be 
a bilinear function of two~$\varphi(\Lambda)$s.
For notational simplicity, we  set $f(\Lambda_1,\,\Lambda_2)\equiv \tilde f( \varphi(\Lambda_1),\,\varphi(\Lambda_2))\,.$
We assume that the function $f(\Lambda_1,\,\Lambda_2)$ is bounded for $0\le \Lambda_1,\,\Lambda_2 <\infty$. 
We also assume that the following limits 
exist:
\begin{align}\notag
\lim_{a\to \infty}\left(\lim_{\Lambda\to\infty}f(\Lambda,\,a\Lambda)\right)
\quad
\text{and}
\quad
\lim_{a\to 0+}\left(\lim_{\Lambda\to\infty}f(\Lambda,\,a\Lambda)\right)
.
\label{condition2}
\end{align}
Under these conditions, we can prove the following identity:
\begin{equation}\label{relation0}
\lim_{\Lambda\to\infty}
\tilde f( \varphi_R(\Lambda),\,\varphi_R(\Lambda) )
=\frac{1}{2}\lim_{a\to \infty}
\left(\lim_{\Lambda\to\infty}
f(\Lambda,\,\,a\Lambda)\right)
+\frac{1}{2}\lim_{a\to 0+}
\left(\lim_{\Lambda\to\infty}
f(\Lambda,\,a\Lambda)\right)\,.
\end{equation}
To prove \eqref{relation0}, we divide the parameter space of 
$(s_1, s_2)\in [0,\ 1]\times [0,\ 1]$ into three parts:
\[S_1\equiv \{(s_1,\,s_2)\big|0\le s_1<s_2\le 1\}\,,\]
\[S_2\equiv \{(s_1,\,s_2)\big|0\le s_2<s_1\le 1\}\,,\]
and
\[L\equiv\{(s_1,\,s_2)\big|0\le s_1,\,s_2\le 1, \ s_1=s_2\}\,.\]
From \eqref{imp}, we see that 
if the parameters $(s_1,\,s_2)$ belong to $S_1$ or $S_2$, 
then
the limit of the function 
$f(\lambda(\Lambda;\,s_1),\,
\lambda(\Lambda;\,s_2))$
as $\Lambda$ approaches $\infty$  
can be expressed as follows: 
\begin{equation}\notag
\begin{split}
\lim_{\Lambda\to\infty}
f(\lambda(\Lambda;\,s_1),\,
\lambda(\Lambda;\,s_2))
&= \lim_{a\to\infty}
\left(\lim_{\Lambda\to\infty} 
f(\Lambda,\,a\Lambda)\right),
\qquad (s_1,\ s_2)\in S_1\,,
\end{split}
\end{equation}
\begin{equation}\notag
\begin{split}
\lim_{\Lambda\to \infty}
f(\lambda(\Lambda;\,s_1),\,
\lambda(\Lambda;\,s_2))
&= \lim_{a\to 0+}
\left(\lim_{\Lambda\to\infty} f(\Lambda,\,a\Lambda)\right),
\qquad (s_1,\ s_2)\in S_2\,.
\end{split}
\end{equation}
Thus, we find that 
\begin{equation}
\begin{split}
&
\lim_{\Lambda\to\infty}
\tilde f( \varphi_R(\Lambda),\,\varphi_R(\Lambda) )
\\
=
&
\lim_{\Lambda\to \infty}
\int_0^1ds_1\int_0^1ds_2
f(\lambda(\Lambda;\, s_1),\,
\lambda(\Lambda;\,s_2))
\\
=
&
\lim_{\Lambda\to \infty}
\iint_{S_1} ds_1ds_2
f(\lambda(\Lambda;\, s_1),\,
\lambda(\Lambda;\,s_2))
+
\lim_{\Lambda\to \infty}
\iint_{S_2} ds_1ds_2
f(\lambda(\Lambda;\, s_1),\,
\lambda(\Lambda;\,s_2))
\\
&
+
\lim_{\Lambda\to \infty}
\iint_{L} ds_1ds_2
f(\lambda(\Lambda;\, s_1),\,
\lambda(\Lambda;\,s_2))
\\
=
&
\frac{1}{2}\lim_{a\to \infty}
\left(\lim_{\Lambda\to\infty}
f(\Lambda,\,\,a\Lambda)\right)
+\frac{1}{2}\lim_{a\to 0+}
\left(\lim_{\Lambda\to\infty}
f(\Lambda,\,a\Lambda)\right)\,.
\end{split}
\end{equation}
In like manner, if $\tilde f_3$ is a 
bounded, trilinear function, we can show that 
\begin{equation}
\begin{split}\label{3var}
\lim_{\Lambda\to\infty}
\tilde f_3( \varphi_R(\Lambda),\,\varphi_R(\Lambda),\, \varphi_R(\Lambda))
=&\frac{1}{3}\lim_{(a,\,b,\,c)\to (1,\,0,\,0)}
\left(\lim_{\Lambda\to\infty}
f_3(a\Lambda,\,\,b\Lambda,\,c\Lambda)\right)\\
&+\frac{1}{3}\lim_{{(a,\,b,\,c)\to (0,\,1,\,0)}}
\left(\lim_{\Lambda\to\infty}
f_3(a\Lambda,\,b\Lambda,\,c\Lambda)\right)\\
&+\frac{1}{3}\lim_{{(a,\,b,\,c)\to (0,\,0,\,1)}}
\left(\lim_{\Lambda\to\infty}
f_3(a\Lambda,\,b\Lambda,\,c\Lambda)\right)\,.
\end{split}
\end{equation}
Here $f_3(\Lambda_1,\, \Lambda_2,\, \Lambda_3)$ denotes $\tilde f(\varphi(\Lambda_1), \varphi(\Lambda_2), \varphi(\Lambda_3))$, 
and we assumed that the limits on the right-hand side of \eqref{3var} exist. 
To be precise, the identities \eqref{relation0}  and \eqref{3var}  hold
under milder conditions; however, we 
shall not pursue this matter here.

\subsection{Regularized definition}

The regularized form of the solution is given as follows:
\begin{equation}\label{c1}
\Psi=-
\lim_{\Lambda\to\infty}
\int_0^1 ds\int_0^{\lambda(\Lambda;\,s)} dx \,e^{xK}
c\frac{K^2B}{K-1}c\,.
\end{equation}
For convenience, we also define 
a string field 
$\Psi_{\rm cutoff}(\Lambda)$ as
\begin{equation}
\Psi_{\rm cutoff}(\Lambda)=
-
\int_0^\Lambda dx \,e^{xK}
c\frac{K^2B}{K-1}c\,.
\end{equation}
Note that 
\[\Psi=\lim_{\Lambda\to\infty}\int_0^1ds\, \Psi_{\rm cutoff}(\lambda(\Lambda;\,s))\,,\]
which corresponds to the expression 
\eqref{regularized_varphi}.

\subsection{Energy density}\label{cutoff_sol1}
In this subsection, we calculate the energy density of the solution \eqref{c1}. 
\subsubsection{Kinetic term}
\label{kinetic}
We start with evaluation of
  the normalized kinetic term $\widehat {\mathcal E}_K(\Psi)$ for the solution 
\eqref{c1},
 defined by
\begin{equation}
\widehat {\mathcal E}_K(\Psi)=\frac{\pi^2}{3}\langle \,\Psi,\,Q\Psi\,\rangle\,.
\end{equation}
Using the correlation function $C_K(x,y;u,v)\equiv{\rm tr}\left[\,e^{xK}ce^{uK}Bc\,Q(e^{yK}ce^{vK}Bc)\,
\right]$, 
we define the quantity $\widehat{\mathcal E}_K(\Lambda_1,\,\Lambda_2)$ as follows:
\begin{equation}\label{uhb}
\begin{split}
\widehat {\mathcal E}_K(\Lambda_1,\,\Lambda_2)
&=\frac{\pi^2}{3}\,
{\rm tr}\,[\,
\Psi_{\rm cutoff}(\Lambda)\,
 Q\,\Psi_{\rm cutoff}({\Lambda^\prime})\,]\\
                                &=\frac{\pi^2}{3} \int_0^\infty du
\int_0^\infty dv\, e^{-u-v} C_K^{(-1,-1,\,2,\,2)}(\Lambda,\,\Lambda^\prime;\,u,\,v)\,,
\end{split}
\end{equation}
where 
\begin{equation}\label{ijn}
C_K^{(-1,-1,2,2)}(x\,,y;\,u,\,v)\equiv\int^x_0 dx^\prime
\int^y_0dy^\prime\,\frac{\partial^2}{\partial u^2}
\frac{\partial^2}{\partial v^2}C_K(x^\prime,y^\prime;u,v)\,.
\end{equation}
From the relation \eqref{relation0}, it follows that
\begin{equation}\label{ppp}
\widehat {\mathcal E}_K=\frac{1}{2}
\lim_{a\to 0+}\left(\lim_{\Lambda\to\infty}
\widehat {\mathcal E}_K(\Lambda,\,a\Lambda)\right)
+\frac{1}{2}
\lim_{a\to \infty}\left(\lim_{\Lambda\to\infty}
\widehat {\mathcal E}_K(\Lambda,\,a\Lambda)\right)\,.
\end{equation}
We can carry out the differentiation with respect to $u$ and $v$ and the 
integration over $x'$ and $y'$ in 
\eqref{ijn} in a straightforward way.  
Since the integration over $u$ and $v$ 
in \eqref{uhb}
is absolutely convergent, we can take the limit $\Lambda \to \infty$ before 
the integration. 
We then find that
\begin{equation}\notag
\begin{split}
\lim_{\Lambda\to\infty}\widehat {\mathcal E}_K
(\Lambda,\,a\Lambda)
=  \frac{1}{3}
\left\{
\left(
1+\frac{2a\pi^2}
{(1+a)^2}
\right)
\cos\left(\frac{2  \pi}{1+a}\right)
-\pi\left(1-\frac{2}{1+a}
\right)
\sin\left(\frac{ 2 \pi}{1+a}\right)+2\right\}.
\end{split}
\end{equation}
Plugging this expression into \eqref{ppp}, we obtain that
\begin{equation}
\widehat{\mathcal E}_K=1\,.
\end{equation}
%

\subsubsection{Cubic term}
Let us move on to the cubic term.  
We define  the regularized cubic term $\widehat {\mathcal E}_C$ as
\begin{equation}
\widehat {\mathcal E}_C(\Psi)=-\frac{\pi^2}{3}\langle
 \Psi,\,\Psi\ast\Psi\rangle\,.
\end{equation}
We also define the quantity 
$\widehat{\mathcal E}_C(\Lambda_1,\,\Lambda_2, \,\Lambda_3)$
as follows:
\begin{equation}
\begin{split}
\widehat{ \mathcal E}_C(
\Lambda_1,\,\Lambda_2,\,\Lambda_3)
=&{\rm tr}\left[
\Psi_{\rm cutoff}(\Lambda_1)\,
\Psi_{\rm cutoff}(\Lambda_2)\,
\Psi_{\rm cutoff}(\Lambda_3)
\right]\\
\equiv &\int_0^{\Lambda_1} dx \int_0^{\Lambda_2} dy \int_0^{\Lambda_3} dz 
\int_0^\infty du\int_0^\infty dv\int_0^\infty dw\\
                                                                    &
\qquad\times e^{-u-v-w}\frac{\partial^2}{\partial u^2}
\frac{\partial^2}{\partial v^2} \frac{\partial^2}{\partial w^2} 
C_C(x,y,z;u,v,w)\,.
\end{split}
\end{equation}
Then, using \eqref{3var}, $\widehat {\mathcal E}_C $ is given by
\begin{equation}\label{relation3}
\begin{split}
\widehat {\mathcal E}_C(\Psi)
=&\frac{\pi^2}{3}\frac{1}{3}
\lim_{(a,\,b,\,c)\to (1,\,0+,\,0+)}\lim_{\Lambda\to \infty}
\widehat{\mathcal E}_C(a\Lambda,\,b\Lambda,\,c\Lambda)\\
&+\frac{\pi^2}{3}\frac{1}{3}
\lim_{(a,\,b,\,c)\to (0+,\,1,\,0+)}\lim_{\Lambda\to \infty}
\widehat{\mathcal E}_C(a\Lambda,\,b\Lambda,\,c\Lambda)\\
&+\frac{\pi^2}{3}\frac{1}{3}
\lim_{(a,\,b,\,c)\to (0+,\,0+,\,1)}\lim_{\Lambda\to \infty}
\widehat{\mathcal E}_C(a\Lambda,\,b\Lambda,\,c\Lambda)\,.
\end{split}
\end{equation}
It is straightforward to derive the following expression:
\begin{equation}\notag
\begin{split}
&\lim_{\Lambda\to\infty}
\widehat {\mathcal E}_C(\Lambda,\,
a \Lambda,\,b\Lambda)\\
=&
1 +\frac{(1+a)^2+2 a\pi^2}{3(1+a)^2}
\cos\left(\frac{2\pi}{1+a}\right)
+\frac{(1+b)^2+2 b\pi^2}{3(1+b)^2}\cos\left(\frac{2\pi}{1+b}\right)\\
                   &+\frac{(a+b)^2+2 ab\pi^2}{3(a+b)^2}\cos\left(\frac{2a \pi}{a+b}\right)\\
                   &+\frac{1}{3(1+a+b)^3}\Bigg(\{-(1+a+b)^2(-1+2a+2b)-2(a+b)^2\pi^2\}
\cos\left(\frac{2\pi}{1+a+b}\right)\\
                   &\qquad\qquad\qquad\qquad +\{-(1+a+b)^2(2-a+2b)-2a(1+b)^2\pi^2\}
\cos\left(\frac{2a\pi}{1+a+b}\right)\\
                   &\qquad\qquad\qquad\qquad +\{-(1+a+b)^2(2-b+2a)-2b(1+a)^2\pi^2\}
\cos\left(\frac{2b\pi}{1+a+b}\right)\Bigg)\\
                   &+ \frac{(1-a)\pi}{3(1+a)}
\sin\left(\frac{2\pi}{1+a}\right)
+\frac{(1-b)\pi}{3(1+b)}
\sin\left(\frac{2\pi}{1+b} \right)\\
                   &+\frac{(a-b)\pi}{3(a+b)}\sin\left(\frac{2a\pi}{a+b}\right)\\
                   &+\frac{1}{6(1+a+b)^2\pi}\Bigg(\{3(1+a+b)^2-2(-2+a+b)(a+b)\pi^2\}
\sin\left(\frac{2\pi}{1+a+b}\right)\\
                   &\qquad\qquad\qquad\qquad +      \{3(1+a+b)^2-2(1+b-2a)(1+b)\pi^2\}
\sin\left(\frac{2a\pi}{1+a+b}\right)\\
                   &\qquad\qquad\qquad\qquad +      \{3(1+a+b)^2-2(1+a-2b)(1+a)\pi^2\}
\sin\left(\frac{2b\pi}{1+a+b}\right)\Bigg)\,.\\
\end{split}
\end{equation}
From this expression, we find that
\begin{equation}
\widehat{\mathcal{E}}_C=1\,.\end{equation}
%

\subsection{Equation of motion contracted with states in the Fock space}
From the calculation in the proceeding subsection, 
we conclude that 
the equation of motion 
is satisfied 
when it is contracted with the solution itself. 
Now, let us study the 
equation of motion contracted with 
states in the Fock space. 
For convenience, we define the string field $\Psi_\Lambda$ as follows: 
\begin{equation}
\Psi_\Lambda\equiv 
-
\int_0^1 ds\int_0^{\lambda(\Lambda;\,s)} dx \,e^{xK}
c\frac{K^2B}{K-1}c\,.
\end{equation}
The remainder of 
the equation of motion, 
$\rm{eom}(\Psi_\Lambda)\equiv Q\Psi_\Lambda+\Psi_{\Lambda}\ast\Psi_{\Lambda}$, is not zero for finite $\Lambda$.  Its explicit form is given by
\begin{equation}
\begin{split}
\rm eom(\Psi_\Lambda)
=&\frac{1}{K_\Lambda}c\frac{K}{K-1}\left(1-K\frac{1}{K_\Lambda}\right)c\frac{K^2}{K-1}Bc
-\frac{1}{K_\Lambda}c\frac{K}{K-1}\left(1-K\frac{1}{K_\Lambda}\right)c\frac{K^2}{K-1}Bc,
\end{split}
\end{equation}
where
\[\frac{1}{K_\Lambda}\equiv -
\int_0^1 ds
\int_0^{\lambda(\Lambda;\, s)} dx e^{xK}\,.\]

We use the $\mathcal L_0$ Fock basis 
instead of the $L_0$ Fock basis.  
The $\mathcal L_0$ Fock basis is obtained by acting finite number of creation operators 
written in $z=({2}/{\pi})\arctan \xi$ coordinates on the vacuum state~$|0\rangle$. 
Let $\tilde c_n$ denotes oscillators 
of the $c$-ghost
in the $z$ coordinates. 
For a few examples, 
\begin{equation}\notag
c_1\,|\,0\,\rangle=\frac{\pi}{2}\tilde c_1\,|\,0\,\rangle\,, 
\end{equation}
and
\begin{equation}\notag
c_0\,|\,0\,\rangle=\left(\frac{\pi}{2}\right)^2\tilde c_0\,|\,0\,\rangle\,. 
\end{equation}
The $\mathcal L_0$ level of the state $\phi$ is defined by its 
 $\mathcal L_0$ eigenvalue plus one. 
We here calculate 
${\rm tr}[\,{\rm eom}(\Psi(\Lambda))\,\phi\,]$ 
for a few $\mathcal L_0$ levels.
Let $\phi_{m,n}$ denotes a state of the form
\begin{equation}
\phi_{n,m}=e^{K/2}K^ncK^me^{K/2}.
\end{equation}
After some calculation, we obtain   
\begin{equation}
\begin{split}
{\rm tr}[\,{\rm eom}(\Psi)\,\phi_{0,\,0}\,]
= &-\frac{1}{2} \left(\lim_{a\to 0}+\lim_{a\to\infty}\right)\lim_{\Lambda\to\infty} 
\int_0^\Lambda dx\int_0^\infty du\int_0^\infty dv\,e^{-u-v}\\
                   &\qquad\qquad
                   \frac{\partial}{\partial u}\frac{\partial^2}{\partial v^2}\left\{ F\left(\frac{1}{2},\,\frac{1}{2}
+x,\,v,\,a\Lambda+u\right)-F\left(\frac{1}{2},\,\frac{1}{2}+x,\,a\Lambda+u,\,v\right) \right\}\\
                  =&-\frac{1}{2}\left(\lim_{a\to 0}+\lim_{a\to\infty}\right)\Bigg(\frac{1}{2\pi^2}
+\frac{1}{1+a}+\frac{-(1+a)^2+2\pi^2}{2(1+a)^2\pi^2}\cos\left(\frac{2a\pi}{1+a}\right)\\
                   &\qquad\qquad\qquad+\frac{3+a}{2(1+a)\pi}\sin\left(\frac{2a\pi}{1+a}\right)
+{\rm Ci}(2\pi)+{\rm Ci}\left(\frac{2a\pi}{1+a}\right)\\
                   &\qquad\qquad\qquad+\log\left(\frac{1+a}{a}\right)\Bigg)\\
                  =&-\frac{1}{2} (2-\gamma +\text{Ci}(2 \pi )-\log (2 \pi )),
\end{split}
\end{equation}
\begin{equation}
\begin{split}
{\rm tr}[\,{\rm eom}(\Psi)\,\phi_{1,\,0}\,]=& -\left(\lim_{a\to 0}
+\lim_{a\to\infty}\right)\lim_{\Lambda\to\infty} 
\int_0^\Lambda dx\int_0^\infty du\int_0^\infty dv\,e^{-u-v}\\
                   &\qquad\frac{\partial}{\partial u}\frac{\partial^2}{\partial v^2}
\left\{ F^{(1,\,0,\,0,\,0)}\left(\frac{1}{2},\,\frac{1}{2}
+x,\,v,\,a\Lambda+u\right)-F^{(1,\,0,\,0,\,0)}\left(\frac{1}{2},\,\frac{1}{2}
+x,\,a\Lambda+u,\,v\right) \right\}\\
                  =&-\left(\lim_{a\to 0}
+\lim_{a\to\infty}\right)\Bigg(\frac{1}{2\pi^2}
+\frac{1}{1+a}+\frac{-(1+a)^2+2\pi^2}{2(1+a)^2\pi^2}\cos\left(\frac{2a\pi}{1+a}\right)\\
                   &\qquad\qquad\qquad+\frac{3+a}{2(1+a)\pi}\sin\left(\frac{2a\pi}{1+a}\right)
+{\rm Ci}(2\pi)+{\rm Ci}\left(\frac{2a\pi}{1+a}\right)\\
                   &\qquad\qquad\qquad+\log\left(\frac{1+a}{a}\right)\Bigg)\\
                  =&-2+\gamma -\text{Ci}(2 \pi )+\log (2 \pi )\,,
\end{split}
\end{equation}
and
\begin{equation}
{\rm tr}[\,{\rm eom}(\Psi)\,\phi_{0,\,1}\,]=0\,.
\end{equation}
Here Ci($x$) denotes the cosine integral function
\[{\rm Ci}(x)=-\int_x^\infty \frac{\cos t}{t}dt\,,\]
 and $\gamma$ denotes the  Euler-Mascheroni constant. 
The function $F(x_1,\,x_2,\,x_3,\,x_4)$ denotes the correlation function 
\[F(x_1,\,x_2,\,x_3,\,x_4)=
{\rm tr}\left[\,Bce^{x_1K}ce^{x_2K}ce^{x_3K}ce^{x_4K}\,\right]\,.
\]
Its explicit form is presented in \eqref{s6}. 
We also used the notation 
\[F^{(1,\,0,\,0,\,0)}(x_1,\,x_2,\,x_3,\,x_4)\equiv\frac{\partial}{\partial x_1}F(x_1,\,x_2,\,x_3,\,x_4)\,.\]
The correspondence between $\phi_{m,n}$ and the states 
in the $\mathcal L_0$ Fock space is given by
\begin{equation}
\tilde c_1|\,0\,\rangle\sim e^{\frac{K}{2}}ce^{\frac{K}{2}}=\phi_{0,\,0}\,,
\end{equation}
\begin{equation}
\mathcal L_{-1}\tilde c_1|\,0\,\rangle\sim e^{\frac{K}{2}}(Kc-cK)e^{\frac{K}{2}}=\phi_{1,0}-\phi_{0,1}\,.
\end{equation}
We also see that the following quantity is zero,
\begin{equation}
\langle\,{\rm eom}(\Psi)\,|\, \mathcal L^{\text{matter}}_{-1}\tilde c_1\,|\,0\,\rangle =0\,,
\end{equation}
since the matter one-point function vanishes. 
We then find that
\begin{eqnarray}\label{anm1}
\langle\,{\rm eom}(\Psi)\,|\,\tilde c_1\,|\,0\,\rangle &=&
\frac{1}{2}(-2+\gamma-{\rm Ci}(2\pi)+\log(2\pi))\qquad(\sim 0.218827)\,,\\
\langle\,{\rm eom}(\Psi)\,|\,\tilde c_0\,|\,0\,\rangle &=&
\frac{1}{2}(-2+\gamma-{\rm Ci}(2\pi)+\log(2\pi))\,.
\end{eqnarray}
Thus, we conclude that the equation of motion is broken 
when it is contracted with some states in the Fock space. 
The constants presented in these expressions can be gathered into a single series,
\begin{equation}
\gamma-{\rm Ci}(2\pi)+\log(2\pi)
=-\frac{1}{2}\sum_{k=1}^{\infty}\frac{(-1)^k(2\pi )^{2k}}{k(2k)!}\,.
\end{equation}

\subsubsection*{Comparison to the results in \cite{Murata:2011ep} }
In \cite{Murata:2011ep},  the remainder of the equation of motion under the $\epsilon$-regularization \cite{Murata:2011ep,Hata:2011ke} was minutely investigated. 
In our notation, the regularization of the ansatz \eqref{old-double-brane-solution} under the $\epsilon$-regularization is written as follows:
 \[\Psi_{\epsilon}=\frac{1}{K-\epsilon}c\frac{(K-\epsilon)^2}{K-\epsilon-1}Bc\,.\]
The remainder of the equation of motion $\rm{eom}(\Psi_\epsilon)$ is given by  
\begin{equation}
{\rm tr}\left[\,{\rm eom}(\Psi_\epsilon)\phi_{n,\,m}\,\right]={\rm tr}\left[\,
\frac{-\epsilon}{K-\epsilon}c\frac{(K-\epsilon)^2}{K-\epsilon-1}c\,\phi_{n,\,m}\,\right]\,.\end{equation}
Using this expression, one can derive 
 the following result:
\begin{equation}\label{anm}
\begin{split}
&\lim_{\epsilon\to 0}\langle\, \text{eom}(\Psi_\epsilon)\,|\tilde c_1|\,0\,\rangle\\
=&-\lim_{\epsilon\to 0}\int_0^\infty dx\int_0^\infty dy\,\epsilon
e^{-\epsilon x-(1+\epsilon)y}
\left(\frac{\partial}{\partial y}-\epsilon\right)^2
{\rm tr}\left[\,e^{\left(x+\frac{1}{2}\right)K}ce^{yK}ce^{\frac{1}{2}K}c\,\right]\\
=&1\,.
\end{split}
\end{equation}
This result is consistent with that obtained in \cite{Murata:2011ep} (see the paragraph including (3.15) in \cite{Murata:2011ep}). 

From \eqref{anm1} and \eqref{anm}, 
we conclude that the remainder of the equation of motion depends on choice of the regularization. 

\section{Notation}
\setcounter{equation}{0}
In this appendix, we 
summarize our notation. 
\subsection{Conventions of the star-product}
In this subsection, we clarify conventions of the star-product.\footnote{
To write this appendix, we consult the following textbook in part:\\ 
N. Ishibashi and K. Murakami, ``String Field Theory -- for a deeper understanding of string theory (Gen no ba no riron -- gen riron no yori fukai rikai no tame ni),'' 
Rinji Bessatsu Suuri Kagaku SGC Raiburari-92,
 Saiensu-sha, (2012) [ISSN0386-8257] (in Japanese).
}
Let $\Phi_1$ and $\Phi_2$ be open string fields. 
When we calculate the star-product $\Phi_1 \ast \Phi_2$, 
we glue the right half ($0\le \sigma \le \pi/2$) of $\Phi_1$ to the left half ($\pi/2\le \sigma \le \pi$) of $\Phi_2$. 
We call this convention {\it the right-handed convention}. 
This convention is convenient when we depict pictures of wedge states  in the sliver coordinates. 

On the other hand, in the original definition of the star-product\cite{Witten:1985cc},  
we glue the left half of $\Phi_1$ to the right half of $\Phi_2$ to calculate 
the star-product of $\Phi_1$ and $\Phi_2$. 
We call this convention {\it the left-handed convention}. 
In order to avoid possible confusion, we here write the star-product in the left-handed convention as 
$(\Phi_1 \ast \Phi_2)_L$. 
Translation from one convention to the other is simple:
\[(\Phi_1 \ast \Phi_2)_L=(-1)^{|\Phi_1||\Phi_2|}
\Phi_2 \ast \Phi_1\] 

Classical solutions in the left-handed convention $\Psi_L$ and that in the right-handed 
convention  $\Psi$ are related as follows:
\begin{equation}\label{aaa}\Psi_L=-\Psi\,.\end{equation}
Then, $\Psi_L$ and $\Psi$ satisfy the equation of motion as follows:
\[Q\Psi_L+(\Psi_L\ast\Psi_L)_L=0\,,\]
\[Q\Psi+\Psi\ast\Psi=0\,.\]

\subsection{Definition of $K$, $B$ and $c$}
The $KBc$ subalgebra is 
originally introduced to represent 
a class of wedge states with operator 
insertions.  
Let $|\,n+1\,\rangle$ denote 
the wedge state of width $n$. 
If $n$ is a natural number, then $|\,n+1\,\rangle$ can be written as follows:
\[|\,n+1\,\rangle=\underbrace{|\,0\,\rangle \ast \dots \ast |\,0\,\rangle}_{n}\,.\]
Using 
$K$, it is expressed as follows:
\[|\,n+1\,\rangle=e^{nK}\,.\]
All the functions of $K$ appearing in this paper are defined as a superposition of wedge states, except for the formal object $1/K$: 
\[f(K)=\int_0^\infty dx\, \widetilde f(x)e^{xK}\,.\]
For example, 
\[\frac{1}{1-K}=\int_0^\infty dx\, e^{-x}e^{xK}\,,\]
and
\[K=\frac{\partial}{\partial x}e^{xK}\Big|_{x\to 0+}
\left(=\int_0^\infty dx\,\delta(x)\frac{\partial}{\partial x}e^{xK}\right)\,.\]
This is a definition of $K$\,. 

To clarify the definition of $K$, $B$ and $c$, it is convenient to 
express them 
using the identity string field~$|\,\mathcal I\,\rangle$:\footnote{
The identity string field $|\,\mathcal I\,\rangle$ is usually denoted by $1$ when we express string fields using $K$, $B$ and $c$, as in \eqref{KBc1}.  }
\[K=\int_{i\infty}^{-i\infty}
\frac{dz}{2\pi i}
T(z)|\,\mathcal I\,\rangle\,, \]
\[B=\int_{i\infty}^{-i\infty}
\frac{dz}{2\pi i}
B(z)|\,\mathcal I\,\rangle\,, \]
and
\[c=c(z=1)|\,\mathcal I\,\rangle\,. \]
The coordinate system $z$ is defined by 
\[z=\frac{2}{\pi}\arctan \xi\,,\]
where $\xi$ represents the usual coordinate system 
on the upper half plane which is used in the radial quantization of the open string. 
Note that $K$, $B$ and $c$ are identity-based string fields, and some 
objects made from them are singular. 
For  example, the value of 
${\rm tr}[\,KcKcKc\,]$ is indefinite.

In the left-handed convention,
definition of basic elements of the $KBc$ subalgebra is different from that in the right-handed convention: 
\[K_{L}=-K\,,
\qquad B_L=-B\,,
\qquad c_L=c\,.\]
Under this definition, $K_L$, $B_L$ and $c_L$ satisfy the same algebraic relations as $K$, $B$ and $c$. 
As an example, let us write the solution \eqref{new-double-brane-solution2} in the left-handed convention: 
\[\Psi=Kc\frac{K}{1-K}Bc
\quad
\Leftrightarrow \quad\Psi_L=c_L\frac{K_L}{1+K_L}B_Lc_LK_L\,. \]

At the end of this subsection, we clarify the overall factors of $K$, $B$ and $c$. 
Our $K$ and $B$  are $\pi/2$ times those of Okawa's original definition\cite{Okawa:2006vm}, which we here write as $K_{\rm Okawa}$ and $B_{\rm Okawa}$, respectively;  
our $c$ is $2/\pi$ times that of the original definition, which we write as $c_{\rm Okawa}$: 
\[K=\frac{\pi}{2}K_{\rm Okawa},\quad B=\frac{\pi}{2}B_{\rm Okawa},\quad c=\frac{2}{\pi}c_{\rm Okawa}\,.\]
\subsection{Correlation functions}
\label{notationcorrelation}
We use the notation tr$[\dots]$ to represent correlation functions in  the $KBc$ subalgebra. 
In our notation,  the four point function is expressed as follows: 
\begin{equation} \label{s6}
\begin{split}
&\text{tr}\left[\,Bce^{x_1K}ce^{x_2K}ce^{x_3K}ce^{x_4K}\,\right]
\equiv 
\left\langle\,\mathcal B_L c(0)\,c(x_1)\,c(x_1+x_2)\,c(x_1+x_2+x_3)\,\right\rangle_{C_{x_1+x_2+x_3+x_4}}\\
=-&\frac{s^2}{4\pi^3}
\Big( \,x_3\sin\frac{2\pi x_1}{s} -(x_2+x_3)\sin\frac{2\pi (x_1+x_2)}{s}+ x_2
\sin\frac{2\pi(x_1+x_2+x_3)}{s} \\
                                                                &+
x_1\sin\frac{2\pi x_3}{s}-(x_1+x_2)\sin\frac{2\pi (x_2+x_3)}{s}
+(x_1+x_2+x_3)\sin\frac{2\pi x_2}{s}   \,\Big)\,,\\
\end{split}
\end{equation}
with
\begin{equation} \label{s7}
s=x_1+x_2+x_3+x_4\,.
\end{equation}
Here $C_{r}$ denotes a semi-infinite cylinder of circumference $r$.  
The character $s$ represents 
the circumference of the cylinder; 
$\mathcal B_L$ denotes a line integral of the $b$-ghost
 $\int_{i\infty}^{-i\infty}dz\, b(z)$. 
In this case, the path of integration is along the line Re$(z)=0-$\,. 
 Note that \eqref{s6} is not smooth at $(x_1,\, x_2,\,x_3,\,x_4)=(0,\,0,\,0,\,0)$.  
 This is a source of the singularity of correlation functions discussed in $\S$ 3.2. 

The following two correlators
are used when we calculate the energy density of classical solutions:
\begin{equation} \label{s21}
\begin{split}
&C_K(x,\,y;\,u,\,v)\equiv{\rm tr}\left[\,e^{xK}ce^{uK}Bc\,Q(e^{yK}ce^{vK}Bc)\,\right]\\
=&\frac{1}{2\pi^2}\Bigg\{-(x+y) s+y(s-x)\cos\frac{2\pi x}{s}+x(s-y)\cos\frac{2\pi y}{s}+uv\cos\frac{2\pi u}{s}+uv\cos\frac{2\pi v}{s}\\
&+(xy-uv)\cos\frac{2\pi(x+v)}{s}+(xy-uv)\cos\frac{2\pi(y+v)}{s}\Bigg\}\\
&+\frac{s}{4\pi^3}\Bigg\{ 2y\sin\frac{2\pi x}{s}+2x\sin\frac{2\pi y}{s}+(s-2v)\sin\frac{2\pi u}{s}+(s-2u)\sin\frac{2\pi v}{s}\\
&+(x-y+u-v)\sin\frac{2\pi(x+v)}{s}+(-x+y+u-v)\sin\frac{2\pi(y+v)}{s} \Bigg\}\,,\\
\end{split}
\end{equation}
with
\begin{equation} \label{s22}
s=x+y+u+v\,,
\end{equation}
and 
\begin{equation} \label{s28}
\begin{split}
&C_C(x,y,z;u,v,w)
\equiv {\rm tr}\left[\,e^{xK}ce^{uK}Bce^{yK}cBe^{vK}ce^{zK}cBe^{wK}c\,\right]\\
                =&\frac{s^2}{4\pi^3}x\left( \sin\frac{2\pi v}{s}-\sin\frac{2\pi (v+y)}{s} -\sin\frac{2\pi (v+z)}{s} +\sin\frac{2\pi (v+y+z)}{s}  \right)\\
                &+\frac{s^2}{4\pi^3}y\left( \sin\frac{2\pi w}{s}-\sin\frac{2\pi (w+z)}{s} -\sin\frac{2\pi (w+x)}{s} +\sin\frac{2\pi (w+z+x)}{s}  \right)\\
                &+\frac{s^2}{4\pi^3}z\left( \sin\frac{2\pi u}{s}-\sin\frac{2\pi (u+x)}{s} -\sin\frac{2\pi (u+y)}{s} -\sin\frac{2\pi (u+x+y)}{s}  \right)\,,
\end{split}
\end{equation}
with
\begin{equation} \label{s22}
s=x+y+z+u+v+w\,.
\end{equation}

\end{document}